\pdfoutput=1
\documentclass[
 aip,
 amsmath,amssymb,
 reprint,%
]{revtex4-1}

\usepackage{graphicx}
\usepackage{dcolumn}
\usepackage{bm}

\usepackage[utf8]{inputenc}
\usepackage[T1]{fontenc}
\usepackage{mathptmx}
\usepackage{etoolbox}

\makeatletter
\def\@email#1#2{%
 \endgroup
 \patchcmd{\titleblock@produce}
  {\frontmatter@RRAPformat}
  {\frontmatter@RRAPformat{\produce@RRAP{*#1\href{mailto:#2}{#2}}}\frontmatter@RRAPformat}
  {}{}
}%
\makeatother
\begin{document}

\preprint{AIP/123-QED}

\title{\textit{Ab initio} determination of thermal conductivity in crystals}
\author{Krzysztof.Parlinski}
\affiliation{Institute of Nuclear Physics, Polish Academy of Sciences,
               Radzikowskiego 152, PL-31342 Krak\'ow, Poland}
\affiliation{Computing for Materials, Krak\'ow, Poland}
\email{Krzysztof.Parlinski@ifj.edu.pl}
\date{04 January 2023}
\begin{abstract}
The calculations of thermal conductivity requires to know anharmonic properties
of the crystal. For this purpose a non-perturbative anharmonic theory is applied,
which do not make use of the potential energy expansion over atomic displacements, 
but instead, runs {\textit ab initio} calculations of Hellmann-Feynman forces for 
atomic patterns of atoms with specific displacements to rebuild the anharmonic 
phonon frequencies, and group velocities. see 
[K.Parlinski, Phys.Rev. B \textbf{98}, 054305 (2018),]  
The Green-Kubo equation for the thermal conductivity needs to know the above 
quantities and the phonon relaxation times, which are related to the 4th-order 
phonon correlation function  expressed in terms of phonon anihilation and 
creation Bose operators. In currect formulation of anharmonic theory the 
relaxation times can be derived as analitical expression.
The Green-Kubo formulae was succesfully applied to find thermal conductivity 
of $Si$ and 
conductivities, related to the phonon and elastic waves, respectivily, were computed.
\end{abstract}
\maketitle
\section{INTRODUCTION}
\par
The understanding of thermal conductivity
in solids is needed for applications of technically relevant materials
to nanofabrication technology, to manufacture electronic devices
for nanoscale demands, to  understand the mechanisms, predict the properties
of solid thermal condunctivities and to be able to run related computations. 
Similarly, handle of thermal conductivity describes partly the behaviour
of thermoelectrics, electron-mediate superconductors and 
thermal conductivity materials, which govern the heat transfer processes 
in the Earth's interior.
\par
The heat transport properties of solids are usually divided into two mechanisms:
First kind is called \textbf{Lattice thermal conductivity} (LTC). It is calculated applying
phonon anharmonicity. The method seems to be rather well known, and in this case
the Green-Kubo  linear-response theory \cite{srivastava}  is mainly used.
There are some variants in formulating this method. In one of them 
the harmonic phonon frequencies, 
the group velocities for phonon modes, and some relaxation times are used.
Another way is to find the input data from  the anharmonic perturbation method,
usually with help of  the triple and quatric order terms
\cite{cowley,amaradudin,michel1,michel2,barron,strauch,esfarjani,tadano,grimvall},
where relaxation time comes from solving the Boltzmann equation \cite{callaway,broido}, generally
using third, or third and fourth order anharmonic terms only.
Next method is to run molecular dynamics  (MD)
\cite{hellman50,hellman37,hellman46,scheffler,wimmer,BinWei}
provided that the potential of the studied system is known.  
The last mentioned approach solves the classical equations of motions 
for the system, tracing particle's
evolution and then collecting the necessary quantities, which are required by 
Green-Kubo formulae. Typically, the Green-Kubo equations describe properly 
the thermal conductivity of solids for LTC, in the interval 
from around room to melting temperature.
 The anharmonic effects alone  
can be also studied applying the stochastic self-consistent harmonic approximation method
\cite{errea34,errea117}, which according to the Gibbs – Bogoliubov variational principle
requires that the true free energy of the system reaches the minimum of 
the functional $\mathcal{F}[\tilde {\rho }]$ with respect to all 
possible trial density matrices $[\tilde {\rho }]$.
\par
The second kind of heat transport will be in this case called \textbf{High thermal conductivity} (HTC), 
for which the complete theory is still under construction.
The HTC typically occurs in simple crystal structures. 
At \textit{low temperature} (below 200 - 300K) HTC materials exhibit usually two, even 
three order of magnitude higher thermal conductivity values 
than the same material at high temperature range only (above 300K).
To this group of crystals belongs:
C (diamond), Si, Ge, AlN, AlP, BAs, BN, BP, BeS, GaN, MgO \cite{slack1AlN}.
Thermal properties of several HTC crystals have been 
measured by Slack \textit{ et.al.} 
\cite{slack1AlN,slackMgO,sichel,hollandSiGe,slack3SiGe}. 
These  HTC materials attracted special attention and called for
relevant theory. In 1964 Glassbrenner and Slack \cite{slack3SiGe}
proposed a mechanism of HTC for silicon Si and Germanium Ge,
based on phenomenological approach \cite{hollandSiGe,slack3SiGe}. 
Later, a similar consideration on \textit{ab initio} level
was published by A.Ward \textit{ et.al.} \cite{broido}.
Recently, Esfarjani \textit{ et.al.}  \cite{esfarjani1}, studying Si, 
have discussed HTC mechanism as arising from large mean free path of phonons,
determined by size of sample.
It was shown that HTC of Si arises 
for more than order of magnitude, if mean free path spans 
from about nanometers to 100 microns.
\par
It is generally accepted that the LTC is totally described by 
the acoustic and optic phonon modes, and therefore the LTC heat transfer is 
described by Green-Kubo formulae, where usually
the relaxation time is found by Boltzmann equation or MD simulation.
In articles \cite{Allen4} and \cite{Allen5} it was shown how the Green 
function of the anharmonic perturbation theory may lead 
to  the typical Lorentzian term, 
with shift and width of anharmonic peak  \cite{Allen5}.
\par 
In the article \cite{baroni} it
was  attempted to decouple the fourth-order correlation function, 
responsible for the relaxation times and needed for thermal conductivity,
using the pairing  Wick's theorem, but finally 
MD runs validated the results for silicon.
\par
In the present article we reformulate the Green-Kubo approach to use so called 
displacement patterns (DPs) of atomic configurations to derive the LTC
directly from phonon dispersion curves created from DP and simultaneously determine
phonon relaxation times. Moreover,  derivation of the relaxation time
from solution of the Boltzmann equation or MD calculations is not needed.
\par
In next sections the method has been extended to handle also the HTC phenomena.
In this case the low frequency and very long elastic waves are used to 
govern the HTC process.
To calculate such  long wavelength states we compute \textit{ab initio} 
the elastic constant tensors with the equilibrium atoms in the supercell 
and for  series of similar supercells with atoms displaced       
from equilibrium positions due to presence of temperature, like in DP. 
From these elastic supercells
one calculates the frequencies and group velocities of the 
elastic waves,  and  apply them to the Green-Kubo expression to find HTC. 
In this case it is obvious that the accounted  wavelength of 
elastic waves could be 
considerably longer than wavelength of ordinary phonons, therefore
one must introduce limit to the longest active wavevector which could be 
accommodated in  the sample size.
The LTC and HTC calculation of Green-Kubo relations, as derived in this paper, 
have the same formal forms. 
\subsection{Anharmonicity}
\par 
The thermal conductivity in solids is determined by anharmonicity of the system,
therefore, one should start from discussion how to handle 
anharmonicity. 
In  present article a procedure, which takes also advantage  of
the \textit{ab initio} calculations considers anharmonic properties  of crystals
within a new non-perturbative approach.
(see Ref.  \cite{parlinski}). It would be much much easier to understand the current article
first looking to  Ref. \cite{parlinski} and glance at the examples presented there.
There, the procedure begins from selecting the supercell of the studied crystal
and calculating the harmonic phonons, using \textsc{Phonon} software 
\cite{parlinski3xx,prlparlinski}.
\par
At equilibrium   every atom of the crystal resides near the 
potential energy minimum.
Displacing an atom from its equilibrium position by a vector 
\textbf{u} one creates so called Hellmann-Feynman (HF) forces 
computed using {\sc VASP} \cite{kresse}, and
acting on the  surrounding atoms, in particular atoms of the supercell. 
\textsc{Phonon} computes in this way the harmonic phonon  frequencies 
$\omega^{(0)}({\bf k},j)$ and eigenvectors $e^{(0)}({\bf k},j)$.
The HF forces could be calculated with the \textit{ab initio} program. 
The same HF forces are used 
to build all force constants and  dynamical matrix elements, which are the essential 
quantities in lattice dynamics theory since more as a century \cite{amaradudin}. 
One should only keep in mind that the used atomic displacements 
amplitudes \textbf{u}
should probe only small interval of the harmonic potential
around the atoms. From these data the mentioned software
calculates harmonic phonon dispersion curves  in the whole Brillouin zone.
In the harmonic calculations the used atomic displacements \textbf{u}  
are small, of order of $0.03-0.04$ \AA, which is close to zero-temperature
phonon vibration.
\par
In this harmonic theory  \cite{parlinski} 
the method uses first the \textit{exact}    wavevectors ${\bf k}$, 
with wavelengths, being commensurate with the supercell size. 
At such   \textit{exact wavevectors}\cite{parlinski3xx,prlparlinski,parlinski} 
the periodic structure of the crystals ensures that
the harmonic frequencies $\omega^{(0)}({\bf k},j)$ 
and eigenvectors $e^{(0)}({\bf k},j)$
are calculated \textit{exactly}, independent on the size of the supercell.
Unfortunately, the list of exact wavevectors  diminishes 
with decreasing size of the supercell. Of course, certain balance between 
computational time and accuracy of the result will  determine the selected 
supercell size.
The  phonon frequencies and eigenvectors 
beyond exact wavevectors
are  interpolations between exact wavevectors. The interpolations are supported by
a traditional analytical derivation of dynamical matrix elements, which must be solved,
 what in practice leads to the valid results in the whole Brillouin zone. 
The interpolated procedure uses 
the singular value decomposition $(SVD)$ method \cite{prlparlinski,svd}, 
which simply assures that the finale
phonon dispersion curves are the best fit in the mean square sens 
to the exact phonons frequencies of the exact points within the constrains of 
classical phonon dispersion curves. 
As a matter of fact 
this approach  \cite{prlparlinski} to phonon theory was already equipped in 1996 
with the  procedure similar to the machine learning method. 
\par
The \textsc{Phonon} software \cite{parlinski}  is also able to 
calculate the phonon dispersion curves from supercell with many atoms, 
which are displaced simultaneously out from their equilibrium positions. 
Moreover, if the atomic displacements stay small, it means 
they do not enter
the non-parabolic part of the  potential, then  
the resulting phonon dispersion curves look like in the harmonic case. 
Indeed, the force constants are determined by proportionality
coefficient between atomic displacement and HF force and  
in harmonic regime do not depend on the amplitude of displacements. 
\par
However, if in the above procedure the displacements are larger,  
some deviation of the phonon frequencies might be observed because in reality 
atoms during vibrations visit the non-parabolic parts of the potential.
These changes of frequencies and eigenvectors manifest the {\bf anharmonicity}.
Hence, the deviation of the particular phonon frequency 
$(\omega^{(anh)}({\bf k},j) - \omega^{(0)}({\bf k},j))$,
for the same $({\bf k},j)$, could be considered as a  measure of the anharmonicity
\par
Of course, it is well known that atoms vibrate in the crystal sites due to 
finite temperature $T$.
For a given $T$, one should  displace the atoms from their  
equilibrium positions and  
create the  {\bf displacements pattern (DP)} next used to find the 
phonon vibrations. 
\par
At a given $T$, the DP could be represented as a snapshot of supercell  
with many atoms displaced. 
One would like to create sets of $N_i$  atomic displacement patterns DP$\,^{(i)}$, 
$i=1,2,\dots N_i$, which might arise in the crystal at a given $T$, 
and in different moments and locations.
The proposition given in \cite{parlinski} is as follows:
Each supercell DP should be filled with the phonon waves, determined by 
the well known expression of atomic displacements ${\bf u}({\bf m},\mu,\gamma)$
and supplemented by the phase factor $\phi({\bf k},j)$ 
of traveled phonon  waves, where meaning of indices is later given before Eq.(\ref{crystalflux}).
\begin{equation}
{\bf u}({\bf m},\mu,\gamma) = 
           \frac{Q({\bf k},j)}{\sqrt{M_\mu}}
         e_{\gamma}({\bf k},j\mid \mu )
          exp[2\pi i({\bf k}\cdot {\bf R}({\bf m}, \mu) - \phi({\bf k},j) ]
\label{wave}
\end{equation}
\begin{figure}[t!]
\hspace*{-1.65cm}\includegraphics[width=0.65\textwidth]{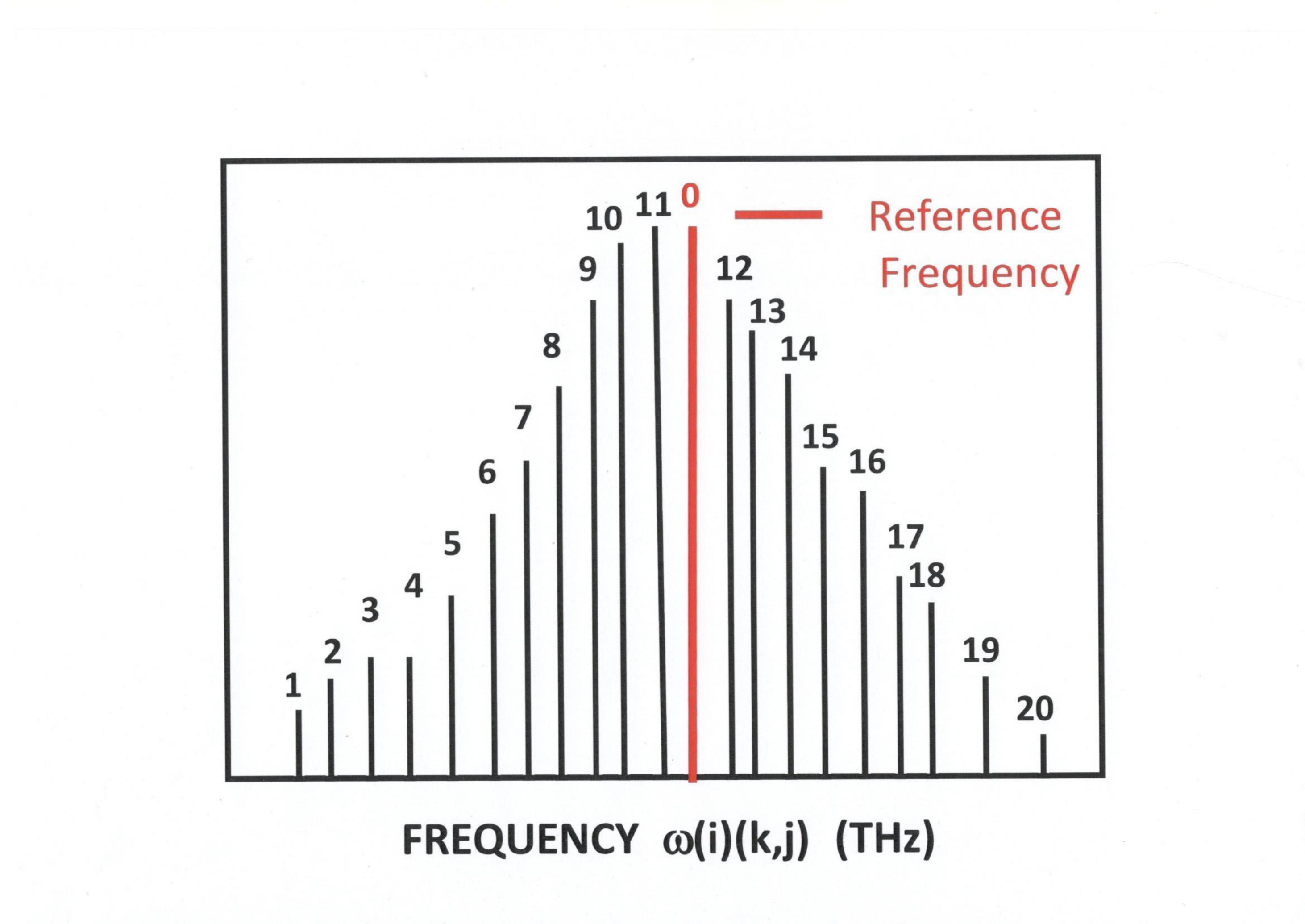} 
\caption {A schematic set of DP$\,^{(i)}$ in single anharmonic phonon peak.}
\label{1DeltaPattern1}
\end{figure} 
The phase $\phi({\bf k},j)$ of the phonon wave could be taken at random from 
the interval $[0.0 - 1.0)$ to mimic different atomic
displacement pattern labelled by the same $({\bf k},j)$. 
The mean square displacement amplitude $<Q^2({\bf k},j)>$ 
of the phonon wave was determined in \cite{debye,ott} by
\begin{figure}[t!]
\includegraphics[width=0.45\textwidth]{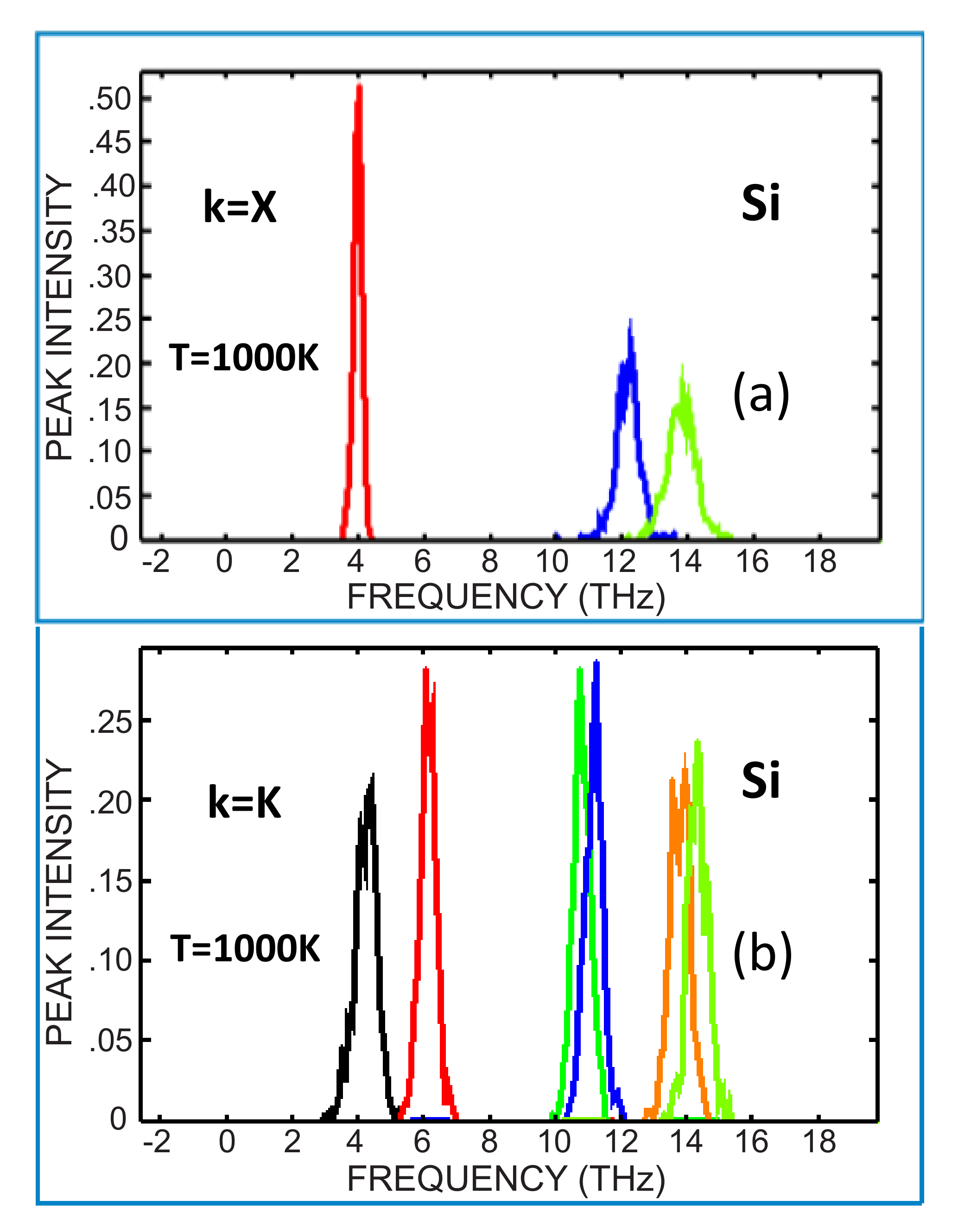}
\caption{Silicon $Si$. Anharmonic phonon peaks calculated for crystal at $T=1000K$
and wavevectors (a) $X = (0.5, 0.5, 0.0$) and (b)  $K = (0.375, 0.375, 0.725)$.
The plots arrived from DP$\,^{(i)}$ $i= 1,2, \dots  500$.}
\label{2FigureDelta2}
\end{figure}  
\begin{equation}
<Q^2({\bf k},j)>  =  
\frac{\hbar  }{2\omega ({\bf k},j)} coth\left(\frac{\hbar \omega({\bf k},j)}
{2k_BT} \right)
\label{amplitude}
\end{equation}
In the harmonic approximation  the above relation is exact.
The Si and MgO the $2 \times 2 \times 2$ supercell contains 64 atoms, 
32 exact wavevectors each with 6 degree of freedom. 
Moreover, the phonon waves may still be supplemented 
by random number of phase $\phi({\bf k},j)$ 
from the interval $[0.0 - 1.0)$.
For Si and MgO, the atomic displacement changes with $T$, from
Eq.(\ref{amplitude}),
it follows $Q=0.05 - 0.16$ \AA $\,\,$ in temperature range  $T=40 - 1500K$. 
This is  only $0.02 - 0.07\%$, respectively, of the nearest 
neighbor interatomic distance.
\begin{figure}[t!]
\includegraphics[width=0.45\textwidth]{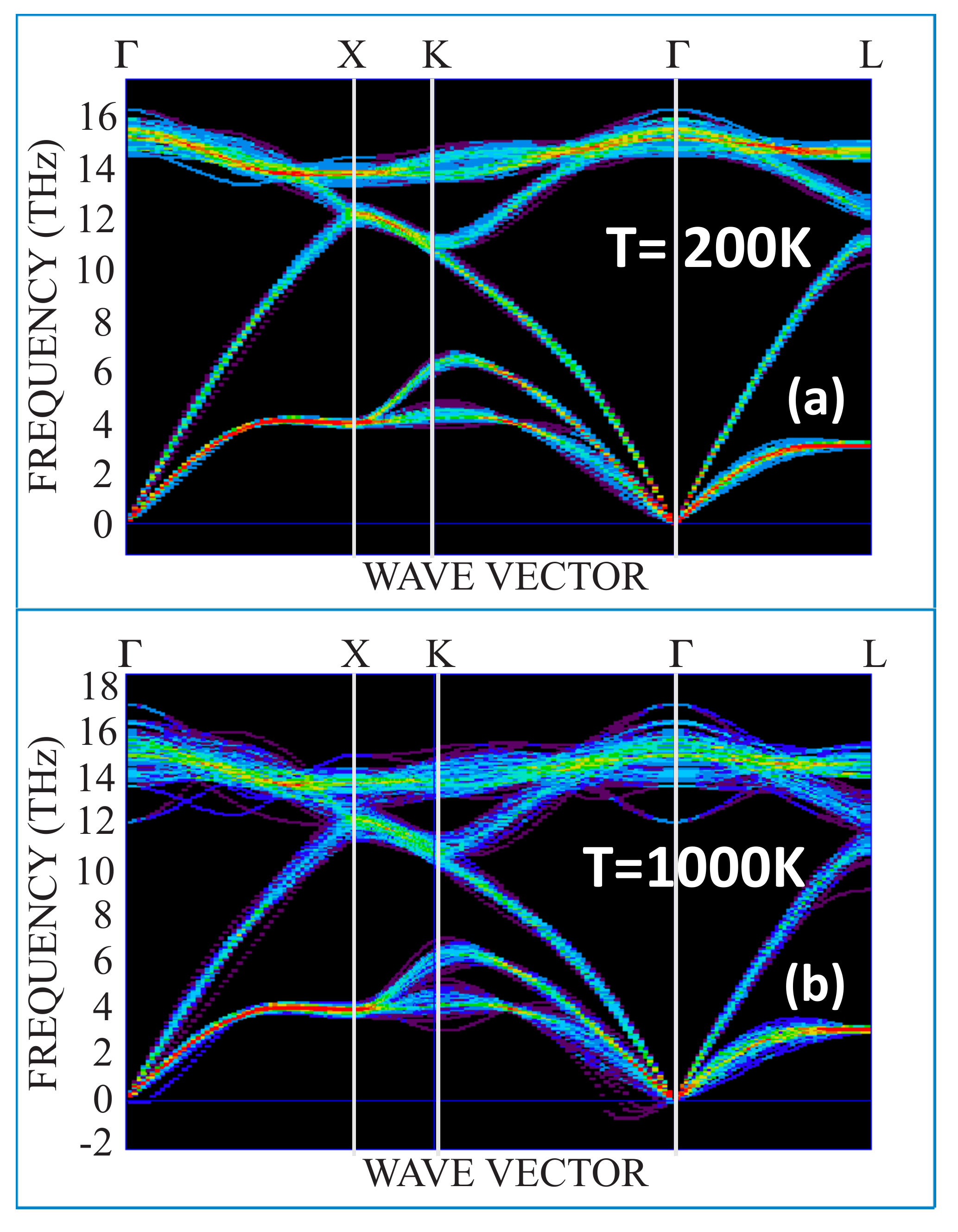}  
\caption{Silicon $Si$. The maps of
anharmonic phonon dispersion curves along 
the line of wavevector $\Gamma - X -K - \Gamma - L$
for temperature (a) $T=200K$ and (b) $T=1000K$
calculated from  DP$\,^{(i)}$ $ 1, 2, \dots  500$ each.
Blue-green-red colours indicates intensity.} 
\label{3FigureMapSi3}  https://www.overleaf.com/project/63668958d1320f3c7c9a7540
\end{figure}
\par 
Using the above method it is rather easy to obtain the anharmonic peaks for any 
wavevector ${\bf k}$ and phonon branch $j$.
These can be any wavevectors, although those which do not belong to list of exact
wavevectors. One needs to create  displacement patterns DP$\,^{(i)}$, 
$i=1,2,\dots N_i$ in the range from $N_i=20$ to $500$, depending on the requested precision. 
For conventional anharmonic peaks 
it could be limited to about $N_i=50$ DP, but to study a peculiar form of the anharmonic peak, 
such as asymmetric shape, particularly high background under the peak of non - Lorentzian shape, 
or even splitting of the single anharmonic peak, the value of $N_i$ should be larger  $N_i=200 - 500$.
The amplitudes of the vibrating atoms causing  anharmonic effects and estimated above 
occur in real crystals and create many HF forces. 
These multiplicity of forces create multiplicity of force constants,
which in turn, are used to solve the equations pf
classical lattice dynamic. 
Schematically the construction of anharmonic phonon mode can be performed
as shown  on Fig.\ref{1DeltaPattern1}. 
Examples of calculated anharmonic phonon peaks 
are shown on Fig.\ref{2FigureDelta2}.
It is a set of $\delta(\omega^{(i)})$ functions of 
 $DP^{(i)}$1,2,$i= \cdots 20$, 
Every segment $i$ represents single snapshot of atomic displacements
for the same  anharmonic phonon mode. The phonon waves have
different phases counting against the fixed sites of the atoms,
hence the frequencies and intensities may vary a little.
The $\delta(\omega^{(i)})$ frequencies together with
intensities (amplitude)
are solutions of the lattice dynamic equations 
for the selected wavevector $\bf k$ and  accompanied displacements
corresponding to temperature $T$.
In the above scheme a 
set of 20 $\delta$s mimic envelope of single anharmonic phonon mode. 
In further one
calculation of anharmonic phonon mode with wavevector ${\bf k}$ 
being located in between the already plotted one 
can be added to increase statistic and precision of phonon peak. 
The envelope of the delta set should give the form of the anharmonic peak.
Reference frequency on the plot corresponds to harmonic frequency
used letter in the conductivity theory.
\par
There appear more profits, following this method. Namely, in this theory 
 {\it the symmetry} 
of each obtained anharmonic peak is uniquely labeled by 
the irreducible representation of the crystal space group.
Normally, it is done only for the harmonic phonon $\delta$-kind peaks.
Here, however,
the calculated area under the anharmonic phonon peaks is characterized 
by the same irreducible representation.
\par 
From the same DP$\,^{(i)}$, $i=1,2,\dots N_i$, with value $N_i$ as discussed above, 
one may construct histograms for the  phonon dispersion curves along any path 
of the reciprocal space, which next can be plotted as a map of the phonon dispersion curves.
Such maps for Si at $T=200K$ and $1000K$ are shown on Fig.\ref{3FigureMapSi3}. 
\subsection{Harmonic and anharmonic hamiltonians}
\label{anharmonic}
\par
The vibrational hamiltonian for a crystal in harmonic approximation 
\cite{amaradudin} can be written as
\begin{eqnarray}
  H^{(0)}  &=& \sum_{{\bf m}, \mu, \gamma}\frac{P^{2,(0)}({\bf m}, \mu, \gamma)}{2M_{\mu}}
\nonumber \\
   &+&   \frac{1}{2} \sum_{{\bf m}, \mu, \gamma} \sum_{{\bf n}, \nu, \delta}
      \Phi^{(0)} ({\bf m}, \mu, \gamma ; {\bf n}, \nu, \delta )
 \nonumber \\
         &\times&  (U^{(0)}({\bf m}, \mu, \gamma)  (U^{(0)}{\bf n}, nu, \delta)
\label{energyham}
\end{eqnarray}
where the harmonic force constants  $\Phi^{(0)}$ have been  calculated from the  
Hellman-Feynman forces of the perfect crystal with atoms  preserving the crystal symmetry.
The $H^{(0)}$ hamiltonian describes the harmonic phonons.
Solving the  eigenvalue equation for $H^{(0)}$ one arrives to harmonic 
phonon frequencies  $\omega^{(0)}({\bf k},j)$ 
and polarization vectors ${\bf e}_{\mu}^{(0)}({\bf k},j)$. 
These collection of harmonic phonons are used 
as a {\it reference set} of  data when analysing the thermal conductivity.
\par 
The current method requires also to find phonon frequencies from the
hamiltonians $H^{(i)}$,where "anharmonic" force constants 
$\Phi^{(i)}$, $i>0$, lead to 
larger/smaller  displacement amplitudes, then in harmonic case. 
Now, one creates the Hellmann-Feynman forces for all displaced atoms collected in 
DP$\,^{(i)}$, Eqs (\ref{wave},\ref{amplitude}).
Solution of these eigenvector equations leads to  little    different phonon frequencies
and one may write 
\begin{eqnarray}
  H^{(i)} &=& \sum_{{\bf m}, \mu, \gamma}\frac{P^{2,(i)}({\bf m}, \mu, \gamma)}{2M_{{\bf m},\mu}}
\nonumber \\
   &+&   \frac{1}{2} \sum_{{\bf m}, \mu, \gamma} \sum_{{\bf n}, \nu, \delta}
      \Phi^{(i)} ( {\bf m}, \mu, \gamma ; {\bf n}, \nu, \delta )
 \nonumber \\
         &\times&  (U^{(i)}({\bf m}, \mu, \gamma)  (U^{(i)}{\bf n}, nu, \delta)
\label{energyanham}
\end{eqnarray}
\par
If the anharmonic system converts to the harmonic one, then the force constants
converge $\Phi^{(i)}$ $\rightarrow$ $\Phi^{(0)}$, and the forces 
are reduced to harmonic one.
From the relations given above we conclude that in similar 
conditions as proclaimed above occurs
$H^{(i)}$ $\rightarrow $ $H^{(0)}$, and therefore 
the anharmonic hamiltonians disappears  $H_{A} = 0$.
Anharmonic hamiltonian vanishes if the phonons of  crystal become harmonic.
Then, the thermal conductivity becomes infinity..
\par
Above, the two body anharmonic force constants, 
$\Phi^{(i)} ( {\bf m}, \mu, \gamma ; {\bf n}, \nu, \delta )$,
are labelled also by index $(i)$ of DP$^{(i)}$,
which indicates that the anharmonic force constant acting on the atom 
$({\bf m}, \mu, \gamma )$  arises not only
due to displacing  a single atom $({\bf n}, \nu, \delta )$
(as was  in the harmonic case), 
but it  really senses also forces coming from all other displaced atoms of supercell 
according to the configuration imposed by  DP$^{(i)}$. 
This suggests that all atoms affects the anharmonic force constant
$\Phi^{(i)} ( {\bf m}, \mu, \gamma ; {\bf n}, \nu, \delta )$ as well.
This means that 
$\Phi^{(i)} ( {\bf m}, \mu, \gamma ; {\bf n}, \nu, \delta )$
is in some sens a many body force constant, which
feels simultaneous displacements of all other atoms in the crystal.
In other words all anharmonic force constants are 
computed not in the perfect crystal, but in the crystal 
being represented by a series of $i= \cdots $, supercells , 
having atoms shifted out from equilibrium positions,
due to finite temperature, 
and from that configuration one computes the contributions to anharmonicity.
\par
The hamiltonian $H^{(0)}$ provides harmonic phonon frequencies only. 
The harmonic potential for perfect insulator should 
lead to infinity thermal conductivity of the crystal.
This statement has been expressed in 
the  textbook of Ashcroft and Mermin \cite{Ashcroft}, 
in Callaway's \cite{callaway} and Maradudin \cite{ltcMaradudin} papers. 
Ashcroft and Mermin says that
 '' in perfect harmonic insulator crystal the phonon scattering does not occur, 
so such a crystal should have infinite thermal conductivity.
Scattering of phonons
from lattice imperfections would produce a finite thermal conductivity, 
but with a  wrong temperature dependence. 
The only way to explain the realistic thermal conductivity data is to 
admit that phonons can
be scattered by other phonons''. Thus, the relevant thermal conductivity
should exhibit the following properties:
(i) demonstrate infinite thermal conductivity for strictly harmonic crystals. 
(ii) describe the finite thermal conductivity for crystal with anharmonicity.
Consequently, one may propose to treat the thermal conductivity using the following approach. 
The anharmonic effects are described by 
the excess of effects arising from   $H^{(i)}$ hamiltonians, 
superimposed on the harmonic modes coming  from $H^{(0)}$. 
Thus, the anharmonicity effects 
of a crystal can be determined  by the following  hamiltonian 
\begin{eqnarray}
H_{A} =  \frac{1}{N_i}\sum_{i=1}^{N_i} \left( H^{(i)} - H^{(0)} \right)
\label{Bdelta} 
\end{eqnarray}
From the relations given above we may conclude  
that for vanishing anharmonicity, when
$H^{(i)}$ $\rightarrow $ $H^{(0)}$, 
the anharmonic hamiltonians
disappear  $H_{A} = 0$ and the crystal exhibits infinite thermal conductivity.
\par
Because the hamiltonians $H^{(0)}$ and 
$H^{(i)}$, Eqs(\ref{energyham},\ref{energyanham}) 
are sums of two positively definite quadratic forms, 
one in the components 
of the momenta and the other in the components
of the atomic displacements, 
it follows from a theorem of matrix algebra \cite{gantmacher}
that it is possible to find  principal axes, or normal 
coordinate transformations 
which simultaneously diagonalized the kinetic 
and potential energies in these hamiltonians.
Such a principal axis transformations are generated 
by the conventional expansion of 
displacements and momenta in terms of plane waves 
and next Bose annihilation 
$b({\bf k},j)$  and creation $b^+({\bf k},j)$ operators.
\par 
In therms of these operators, the hamiltonians Eqs(\ref{energyham},\ref{energyanham}) 
take the simple  forms 
\begin{eqnarray}
  &H^{(0)}  = \sum_{{\bf k},j} \hbar \omega^{(0)}({\bf k},j) [b^+({\bf k},j) b({\bf k},j) + \frac{1}{2}]
 \nonumber \\
  &H^{(i)}  = \sum_{{\bf k},j} \hbar \omega^{(i)}({\bf k},j) [b^+({\bf k},j) b({\bf k},j) + \frac{1}{2}]
\label{hamilanham}
\end{eqnarray}
\par
From Eqs (\ref{Bdelta}, \ref{hamilanham}) the anharmonic hamiltonian $H_{A}$, 
with subtructed harmonic phonon contribution 
$H_{0}$ reads 
\begin{eqnarray}
H_{A} =  \sum_{{\bf k},j} \sum_{i=0}^{N_i}  
    \left(
    \hbar \omega^{(i)}({\bf k},j) - \hbar \omega^{(0)}({\bf k},j) \right) 
\nonumber \\
\times \,\, b^+({\bf k},j) b({\bf k},j)
\label{tothami}
\end{eqnarray}
where it has been assumed that the Bose operators $ b^+({\bf k},j)$ and $b({\bf k},j)$
for the same mode $({\bf k},j)$ with close frequencies should be,
respectively, very similar and further 
we assume that they  remain the same. 
Indeed, in this approach the anharmonicity  is determined by the differences of
$\left( \hbar \omega^{(i)}({\bf k},j) - \hbar \omega^{(0)}({\bf k},j)\right)$.
\par
These frequencies 
could be systematized and collected  to  histograms,
labeled by a wavevector and phonon branch $({\bf k},j$) 
and finally to present as a Lorenzian-kind anharmonic peaks.
Such peaks could be measured by inelastic neutron scattering, Raman spectra, 
or infrared absorption. Below we shall use this method to 
model the thermal conductivity as well.
It is essential to remind that the path from the DP$\,^{(i)}$ to phonon frequencies is 
performed by the solution of lattice dynamics equation of motion only.
\par  
Here, a single DP$\,^{(i)}$ for fixed $i$ can 
be treated as an anharmonic perturbation cluster, arising from
simultaneously displacements of many atoms.
In traditional perturbation theory, DP$\,^{(i)}$ is typically limited to triple or 
quatric interactions. Here, a crystal with supercell of 64 atoms provides 
single  DP$\,^{(i)}$  data for all wavevectors $({\bf k},j)$
of the Brillouin zone, so some cross interaction therms are included.
\section {FORMULAE FOR THERMAL CONDUCTIVITY}
\subsection{Phonons}
\par
The Green-Kubo approach is based on statistical thermodynamics
\cite{green,kubo,kubo2,hardy}.
A derivation of basic formulae  can be found in references
\cite{zwanzig,srivastavabook,kaviany,esfarjani,esfarjani1}.
The heat flux $J(t)$, for simplicity, is usually determined without contribution from 
diffusion and convection, (see Ref. \cite{esfarjani}). 
Here also, we adapt the formalism of the anharmonic theory described in
previous section, to apply the set of anharmonic hamiltonians $H^{(i)}$ Eq.(\ref{Bdelta}). 
The mentioned  method  expects the crystal  to be
presented as a set of $N_i$ supercell's subsystems with atoms randomly displaced 
patterns DP$\,^{(i)}$, $i=1,2,\dots N_i$, corresponding to studied temperature $T$,
\begin{eqnarray}
J_{\alpha} ^{(i)}(t) = \frac{1}{2} \sum_{{\bf m}, \mu, \gamma} \sum_{{\bf n}, \nu, \delta}
(R^{\alpha}({\bf m}, \mu, \gamma) - R^{\alpha}({\bf n}, \nu, \delta))
\nonumber \\
\times  \,\,\,\,
\left( U^{(i)}({\bf m}, \mu, \gamma \mid t) 
\cdot \overline\Phi^{(i)} ({\bf m}, \mu, \gamma ; {\bf n}, \nu, \delta)\right. 
\nonumber \\
\left. \cdot \frac{1}{M_{\nu}}  P^{(i)}({\bf n}, \nu, \delta \mid t)\right)
\nonumber \\
\label{crystalflux}
\end{eqnarray} 
Here, we use indexing of atoms:
first atom: $({\bf m}, \mu, \gamma)$, 
second atom:$({\bf n}, \nu, \delta)$, where 
${\bf m},{\bf n}$ are coordinates of primitive unit cells, 
$\mu,\nu$ are atomic indices within primitive unit cells, and
$\gamma, \delta$ stay for coordinate $x, y, z$.
The force constants  $\Phi^{(i)} ({\bf m}, \mu, \gamma ; {\bf n}, \nu, \delta)$
may have contributions 
from harmonic and/or anharmonic regions of the interatomic potentials. In this sens
the force constants may contain contributions from
any higher order anharmonic therms. Moreover, the force constants might also 
have contributions from other displaced atoms of used DP$\,^{(i)}$, and not shown
explicitly in the now discussed form of $\Phi^{(i)}$. 
The same force constant may also 
represent harmonic force constants.
\par
As argued in Sec.\ref{anharmonic} the thermal conductivity 
should be calculated  according to Eq.(\ref{Thermaltensor}), 
over thermal fluctuations represented by the harmonic 
and anharmonic hamiltonians Eq.(\ref{Bdelta}), determined by the components
DP$\,^{(i)}$ $(i=1,\cdots N_i)$, all generated for the same $T$. 
The Green Kubo expression is then written as
\begin{equation}
\kappa_{\alpha,\beta} = \frac{1}{Vk_BT^2}
\frac{1}{N_{i}} \sum_{i=1}^{N_i}
\int _0^\infty <J^{(i)}_{\alpha}(t) J^{(i)}_{\beta}(0)> dt
\label{Thermaltensor}
\end{equation}
Averaging the above correlation function over DP$\,^{(i)}$ 
one may use it to study also anharmonic phonon peaks. 
Using the expansions of atom displacements and momenta 
over plane waves $Q^{(i)}({\bf k},j)$ and $\overset{\bullet}Q^{(i)}({\bf k},j)$, respectively,
\cite{amaradudin}, one has
\begin{eqnarray}
U^{(i)}({\bf m}, \mu, \gamma\mid t) = \sqrt{\frac{\hbar}{NM_{\mu}}} 
\sum_{\bf k,j} e^{(i)}_\gamma({\bf k},j\mid \mu)
\nonumber \\
\times \,\,\,\, exp[2\pi {\bf i}({\bf k}\cdot {\bf R}({\bf m}, \mu)]
Q^{(i)}({\bf k},j\mid t)
\nonumber \\
P^{(i)}({\bf n}, \nu, \delta \mid t) = \frac{1}{{\bf i}} \sqrt{\frac{\hbar M_\nu}{N}} 
\sum_{\bf k,j} e^{(i)}_\delta({\bf k},j\mid \nu)
\nonumber \\
\times \,\,\,\, exp[2\pi {\bf i}({\bf k}\cdot {\bf R}({\bf n}, \nu)]
\overset{\bullet}{Q}^{(i)} ({\bf k},j \mid t),
\label{normalmodes}
\end{eqnarray} 
where (bold ${\bf i} = \sqrt{-1}$), $N$ is the number of wavevectors ${\bf k}$ used in the summation 
of Eqs (\ref{normalmodes}), and $j$ is the index of phonon branches.
Now, recalculating  Eq.(\ref{crystalflux}) one can rewrite it in the form
\begin{eqnarray}
J^{(i)}_{\alpha}(t) = \frac{\hbar}{{\bf i}N} \sum _{{\bf k},j} \omega^{(i)}({\bf k},j)
 {\bf v}^{(i)\alpha}_{gr}({\bf k},j) 
\nonumber \\ 
\times Q^{(i)}({\bf k},j\mid t) \overset{\bullet}{Q}^{(i)}({\bf k},j \mid t)
\label{jalphaflux}
\end{eqnarray} 
Here, imaginary unit $i$  appears since it was added to 
the exponent of the dynamical matrix  $D^{(i)}({\bf k})$, when used to 
define the group velocity, Eq(\ref{groupvelocity})
\par
In next steps one finds the phonon frequencies and eigenvectors
for perfect crystal $(i=0)$ and for crystal modified with DP$\,{(i)}$, $(i>0)$.
Both are lattice dynamic solutions of the eigenvalue phonon equation
\begin{equation}
\omega^{(i)2}({\bf k},j) = 
{\bf e}^{(i)T}({\bf k},j)
{\bf D}^{(i)}({\bf k})
\cdot {\bf e}^{(i)}({\bf k},j)
\label{eigenvalue}
\end{equation}
Of course they need different values of the elements 
of dynamical matrix  $D^{(i)}({\bf k})$.
\par
Further, the group velocity vectors can be found from relevant dynamical matrices using
\begin{eqnarray}
{\bf v}^{(i)}_{gr}({\bf k},j) &=&  
\frac{1}{2\omega^{(i)}({\bf k},j)  } 
\nonumber \\
&\,&
 \left[{\bf e}^{(i)T}({\bf k},j) \left(
\frac{\partial }{\partial {\bf k}} {\bf D}^{(i)}({\bf k})\right)
\cdot {\bf e}^{(i)}{\bf k},j) \right]
\nonumber \\
\label{groupvelocity}
\end{eqnarray}
Notice that with the same equations the phonon frequencies 
$\omega^{(i)}({\bf k},j)$ and group velocities ${\bf v}^{(i)}_{gr}({\bf k},j)$ 
have been found in \textit{ab initio} procedure via the Hellman-Feynman force
\cite{parlinski} created by displacement of atoms fixed already in DP$\,^{(i)}$'s.
These deviations of DP$\,^{(i)}$ phonon frequencies from the relevant harmonic frequency
contain information concerning the anharmonicity, in terms of frequency and eigenvectors.
\par
Collecting the expressions of Eqs 
(\ref{Thermaltensor}, \ref{jalphaflux}, \ref{groupvelocity})
the thermal conductivity tensor reads
\begin{widetext}
\begin{eqnarray}  
\kappa_{\alpha,\beta}^{LTC} &=& \frac{\hbar^2}{V_{puc}k_BT^2}\frac{1}{N_{i}} \sum_{i=1}^{N_i}
\int _0^\infty dt \frac{1}{Nr}  < \sum_{{\bf k},j} (\omega^{(i)}({\bf k},j))^2
{\bf v}^{(i)\alpha}_{gr}({\bf k},j) {\bf v}^{(i)\beta}_{gr}({\bf k},j) 
\nonumber \\
&\times&
 <Q^{(i)}({\bf k},j\mid t) \overset{\bullet}{Q^{(i)}}({\bf k},j \mid t)
 \overset{\bullet}{Q^{(i)}}({\bf k},j \mid 0) Q^{(i)}({\bf k},j\mid 0>
\label{ThermalConductensornn}
\end{eqnarray}
\end{widetext}
where $r$ is a number of atoms in primitive unit cell, 
${V_{puc}}$ volume of primitive unit cell.
The  appeared fourth  order  phonon  correlation  function 
$<Q(t)\overset{\bullet}{Q}(t) \overset{\bullet}{Q}(0)Q(0)>$
needs some comments.  
It is the only function under the Laplace integral, which depends on time $t$.
If the integrated function would be  a constant $C=const \ne 0$ then
the Laplace integral $\int _0^\infty C dt = \infty$. This would be the  
mechanism to make a  harmonic crystal having infinite thermal conductivity.
\par
To considered the value of the fourth-order correlation function we need 
to express the normal mode amplitudes of phonons by Bose 
annihilation $b$ and creation $b^+$  operators
\begin{eqnarray}
Q^{(i)}({\bf k},j) &=& \frac{1}{\sqrt{2}}
\frac{1}{\sqrt{\omega^{(i)}({\bf k},j)}}
\left( b({\bf k},j) + b^+({\bf -k},j)\right)
\nonumber \\
\overset{\bullet}{Q^{(i)}}({\bf k},j) &=&  \frac{1}{\sqrt{2}}
{\sqrt{\omega^{(i)} ({\bf k},j)}}\left( b({\bf k},j) - b^+({\bf -k},j)\right)
\label{transformation}
\end{eqnarray}
Now, the pair time-dependent correlation functions of $b$, $b^+$ are 
found from the solution of the Heisenberg time dependent 
equations \cite{amaradudin}, in which the 
anharmonic hamiltonian $H_{A}$ Eq.(\ref{Bdelta}) and (\ref{tothami}) has been used. Then
\begin{eqnarray}
<b({\bf k},j\mid t) b^+({\bf k'},j')\mid 0> &=&
\nonumber \\       
     \exp{(-{\bf i}\omega^{(i)}({\bf k},j) - \omega^{(0)}({\bf k},j)] t )}
     (&n^{(i)} &({\bf k},j)+1)
     \delta_{{\bf k,k'}}\delta_{j,j'}
\nonumber \\
<b^+({\bf k},j\mid t) b({\bf k'},j')\mid 0>  &=& 
\nonumber \\
      \exp{(+{\bf i}\omega^{(i)}({\bf k},j) - \omega^{(0)}({\bf k},j)]t )}
      &n^{(i)}&^({\bf k},j) \delta_{{\bf k,k'}}\delta_{j,j'} 
\label{timebose}
\end{eqnarray}
Here, the mean number of phonons in the vibrational mode $({\bf k},j)$ of DP$\,^{(i)}$ at 
temperature $T$, is
\begin{equation}
n^{(i)}({\bf k},j) =\frac{1}{e^{\beta \hbar\omega^{(i)} ({\bf k},j)} - 1}
\label{statistic}
\end{equation}
and $\beta = \frac{1}{k_BT}$. 
\par
Applying Eqs (\ref{transformation}, \ref{timebose}),
the fourth order correlation function 
$<Q(t)\overset{\bullet}{Q}(t) \overset{\bullet}{Q}(0) Q(0)>$ 
can be evaluated, with the Wick'pairing  theorem \cite{baroni,barron}, 
to  16 correlation functions of products of  averages  
consisting of four $b$, $b^+$
operators each. In 10 functions, out of the mentioned 16, 
the  4 operator terms vanish
due to averages  build from pairs of the same kind of operators.
The remaining 6 correlation functions do not vanish 
from the mentioned reasons. 
However, 4 functions arrived  from the last  6 non-zero terms, mutually cancel,
due to averages  build from pairs constructed from the same 
kind of operators and 6 terms
are not vanishing from these reasons. 
However, the 2 last terms
 $<b(t) b(t) b^+(0) b^+(0)>$ and 
 $<b^+(t) b^+(t) b(0) b(0)>$ remain non-zero  and can be written as
\begin{eqnarray}
 <b({\bf k},j\mid t) b({\bf k},j\mid t) b^+({\bf k},&j&\mid 0) b^+({\bf k},j\mid 0)> =
\nonumber \\
          2(n^{(i)}({\bf k},j) +1)^2\,\,\,&e&^{{-2{\bf i}}{[\omega^{(i)}({\bf k},j) - \omega^{(0)}({\bf k},j)}t]} 
\nonumber \\
 <b^+({\bf k},j\mid t) b^+({\bf k},j\mid t) b({\bf k},&j&\mid 0) b({\bf k},j\mid 0)> =
\nonumber \\
          2(n^{(i)}({\bf k},j))^2\,\,\,   &e&^{{+2{\bf i}}{[\omega^{(i)}({\bf k},j) - \omega^{(0)}({\bf k},j)}t]} 
\nonumber \\
\label{bbwzor}
\end{eqnarray}
Applying the time dependence of the surviving pairs in Eq.(\ref{timebose}), 
the non-zero fourth-order correlation functions are 
\begin{eqnarray}
<Q^{(i)}({\bf k},j\mid t) \overset{\bullet}{Q^{(i)}}&(&{\bf k},j \mid t)
 \overset{\bullet}{Q^{(i)}}({\bf k},j \mid 0) Q^{(i)}({\bf k},j\mid 0> =
\nonumber \\
&(&n^{(i)}({\bf k},j)+1)n^{(i)}({\bf k},j) + 1/2) \times
\nonumber \\
\{&(&cos2(\omega^{(i)}({\bf k},j) - \omega^{(0)}({\bf k},j))t\}   
\nonumber \\
- i &(&n^{(i)}({\bf k,j}) + 1/2) \times
\nonumber \\
\{&(&sin2(\omega^{(i)}({\bf k},j) - \omega^{(0)}({\bf k},j))t\}
\label{4pointfunction}
\end{eqnarray}
The above correlation function shows real and imaginary components. 
The time dependence of the real one is governed by the cosine functions,
which always have a maximum at $t=0$. At increased time $t$ 
the integrated functions, being the sum of many
cosines  with  different periods will shrink to a bundle, which by increasing $t$ 
finally converges to zero.
Moreover, one may neglect  $1/2$  because its value appears to be negligible
in comparison to $(n+1) n$ in ranges of typical studied temperature.
\par   
The imaginary term contains periodic sine functions, which start 
from zero at $t=0$. 
At finite values of $t$ the sum of many
sines with  different signs and  periods will lead the 
function shrinking around zero axis to zero bundle, making its 
contribution small. Consequently, we  
neglect the imaginary term as well.
\par
Then, the final thermal factor appears to be equal to
$(n^{(i)}({\bf k},j)+1)n^{(i)}({\bf k},j)$, being the standard form
occurring in the thermal conductivity expressions. 
It is the  occupation coefficient
responsible  for the  thermal distribution. 
Finally, the fourth-order phonon correlation function
reads                    
\begin{eqnarray}
<Q^{(i)}({\bf k},j\mid t) \overset{\bullet}{Q^{(i)}}({\bf k},j \mid t)
 \overset{\bullet}{Q^{(i)}}({\bf k},j \mid 0) Q^{(i)}({\bf k},j\mid 0> =
\nonumber \\
\frac{1}{2} \left((n^{(i)}({\bf k},j)+1)  n^{(i)}({\bf k},j) 
\{cos2(\omega^{(i)}({\bf k},j) - \omega^{(0)}({\bf k},j))t\} \right)
\nonumber \\
\label{4pointrelaxation}
\end{eqnarray}
where the time dependence appears in a cosine function only.
Then, the cosine argument consists of difference of two phonon frequencies.
The first one come from DP$\,^{(i)}$, which is the  
partial information of properties of the 
width of anharmonic peak $({\bf k},j)$ in form of phonon frequency. 
The second is the reference frequency
of the harmonic phonon from the same harmonic mode $\omega^{(0)}({\bf k},j)$.
Below, we  discuss the procedure to derive analytically   the relaxation times 
for the thermal conductivity of anharmonic 
crystals directly from anharmonic theory \cite{parlinski}.
\par
Let us insert the fourth-orders correlation function 
Eq.(\ref{4pointrelaxation})
into relation of the thermal conductivity 
Eq.(\ref{ThermalConductensornn}).
Within the current non-perturbative anharmonic approach \cite{parlinski}
this would be the finale form of the  general Green Kubo relation 
for thermal conductivity in crystals.
\begin{widetext}
\begin{eqnarray}
\kappa_{\alpha,\beta}^{LTC} &=& \frac{\hbar ^2}{NrV_{puc}k_BT^2}
\frac{1}{N_{i}} \sum_{i=1}^{N_i}   \sum_{{\bf k},j} (\omega^{(i)}({\bf k},j))^2
{\bf v}^{(i)\alpha}_{gr}({\bf k},j) {\bf v}^{(i)\beta}_{gr}({\bf k},j) 
\times
(n^{(i)}({\bf k},j) +1)) (n^{(i)}({\bf k},j) 
\times \frac{1}{2} cos[2(\omega^{(i)}({\bf k},j) - \omega^{(0)}({\bf k},j))t]
\label{Thermaltensor2}
\nonumber \\
\end{eqnarray}
\end{widetext}
\subsection{Relaxation Times of Phonon Modes}
\par 
 The last term of Eq.(\ref{Thermaltensor2}) is related to the relaxation function of the DP${\,^{(i)}}$
\begin{equation}
  \tau_0^{(i)}({\bf k},j\mid t) = \frac{1}{2} cos[2(\omega^{(i)}({\bf k},j) - \omega^{(0)}({\bf k},j))t]
\end{equation}
After time-dependent Laplace integration one obtains a partial relaxation times (PRT) 
\begin{equation}
 \tau^{(i)}({\bf k},j) =
   \int _0^\infty  \tau_0^{(i)}({\bf k},j\mid t) \,\,  dt 
\label{Upsilon}
\end{equation}
 Above partial relaxation time is labeled by
 phonon wavevector {\bf k}, phonon branch {\it j} and 
 index $(i)$ of DP$\,^{(i)}$.
 \par
 The conventional relaxation time (CRT)), characterises 
 the complete  $({\bf k},j)$ anharmonic phonon mode being 
 a result of an average of 
 all contributions from DP$\,^{(i)}$. It reads
\begin{eqnarray}
\tau_{con}({\bf k},j) &=&  \frac{1}{N_i} \sum_{i=1}^{N_i} \tau^{(i)}({\bf k},j) = 
\nonumber \\
 &=& \frac{1}{2 N_i} \sum_{i=1}^{N_i}
\int _0^\infty dt \,\, cos(2(\omega^{(i)}({\bf k},j)-\omega^{(0}({\bf k},j))t)
\nonumber \\
\label{conventionalrelaxation}
\end{eqnarray}
This is the relaxation time, which generally is used to
calculate the thermal conductivity.The inverse of CRT is related to the width 
of the phonon anharmonic mode
in frequency space.
\par
To present the computed results of thermal conductivity
we selected out from general relations of Eq.(\ref{Thermaltensor2}), 
two auxiliary  expressions (i)  the time-independent amplitude function,
\begin{eqnarray}
Z_{\alpha,\beta}^{(i)}({\bf k},j)  &=& \left(\frac{k_B}{V_{puc}}\right)  
\left(\frac{\hbar \omega ^{(i)}({\bf k},j)}{k_BT}\right)^2 \times
 \nonumber \\
{\bf v}^{(i)\alpha}_{gr}(&{\bf k},j)& {\bf v}^{(i)\beta}_{gr}({\bf k},j)
(n^{(i)}({\bf k},j) +1) n^{(i)}({\bf k},j)                                                                           \label{Z-function}
\end{eqnarray}
and (ii)  time-dependent Kubo-Green function being the tensor of thermal
conductivity function. This Kubo-Green functions have been plotted on many Figures
of this article. Notice, that here  averaging over index $(i)$, have not yet been applied.)
\begin{equation}
G_{\alpha,\beta}^{(i)}(t) =  \frac{1}{Nr}  \sum_{{\bf k},j}
Z_{\alpha,\beta}^{(i)}({\bf k},j) \tau^{(i)}_0({\bf k},j\mid t)
\label{GKtwo} 
\end{equation}
\par
The above quantity   averaged over DP$\,^{(i)}$ 
leads to averaged Green-Kubo components of the thermal conductivity tensor
\begin{equation}
\kappa_{\alpha,\beta}(t) = \frac{1}{N_i} \sum _{i=1}^{N_i}  \,G^{(i)}_{\alpha,\beta}(t)
\label{GKtime}
\end{equation}
\par
The Laplace integral over components of thermal conductivity tensor,
 Eq.(\ref{GKtime}) gives 
 final tensor of the thermal conductivity. These quantity are given in Tables 
\begin{equation}
\kappa_{\alpha,\beta}  = \int _0^\infty dt \, \kappa_{\alpha,\beta}(t)
\label{GKLaplace}
\end{equation}
\par
Finally, we define the global relaxation time (GRT). It is  
a common average relaxation  time to characterize the 
whole studied system. It includes the Laplace integration,
summation over all DP$\,^{(i)}$) and phonons $({\bf k},j)$
\begin{eqnarray}
\tau_{gl} &=&
\frac{1}{2 N_iNr} \sum_{i=1}^{N_i}
\sum_{{\bf k},j}
\int _0^\infty dt \,\, cos(2(\omega^{(i)}({\bf k},j)-\omega^{(0}({\bf k},j))t) 
\nonumber \\
\label{globalrelaxation}
\end{eqnarray}
\section{LATTICE THERMAL CONDUCTIVITY}
\subsection{Displacement patterns for LTC}
\par
The above described theory will be applied to 
silicon $Si$ and magnesium oxide MgO. These crystals belong to cubic structure with
space groups $Fd\overline{3}m$ and  $Fm\overline{3}m$, respectively,
and each with $r=2$ atoms per primitive unit cells.
All the \textit{ab initio} calculations have been performed using VASP software
\cite{vasp} on $2\times 2\times 2$ supercells with periodic boundary 
conditions and $64$ atoms. 
Using software  {\sc PhononA} \cite{parlinski713}, and calculating the Hellmann-Feynman 
forces created by {\sc VASP} \cite{vasp} the  harmonic phonon dispersion curves
$\omega^{(0)} =\omega^{(0)}({\mathbf k},j)$,  were established and plotted. 
Next, these harmonic curves
were used to create  $N_i$ number of DP$\,^{(i)}$ configurations with additionally 
randomly displaced atoms. All these DP have been used to create sets of 
the Hellman-Feynman forces, specific for each DP, 
created by relaxing the single electronic loop on {\sc VASP}.
\begin{figure}[t!]
\includegraphics[width=0.45\textwidth]{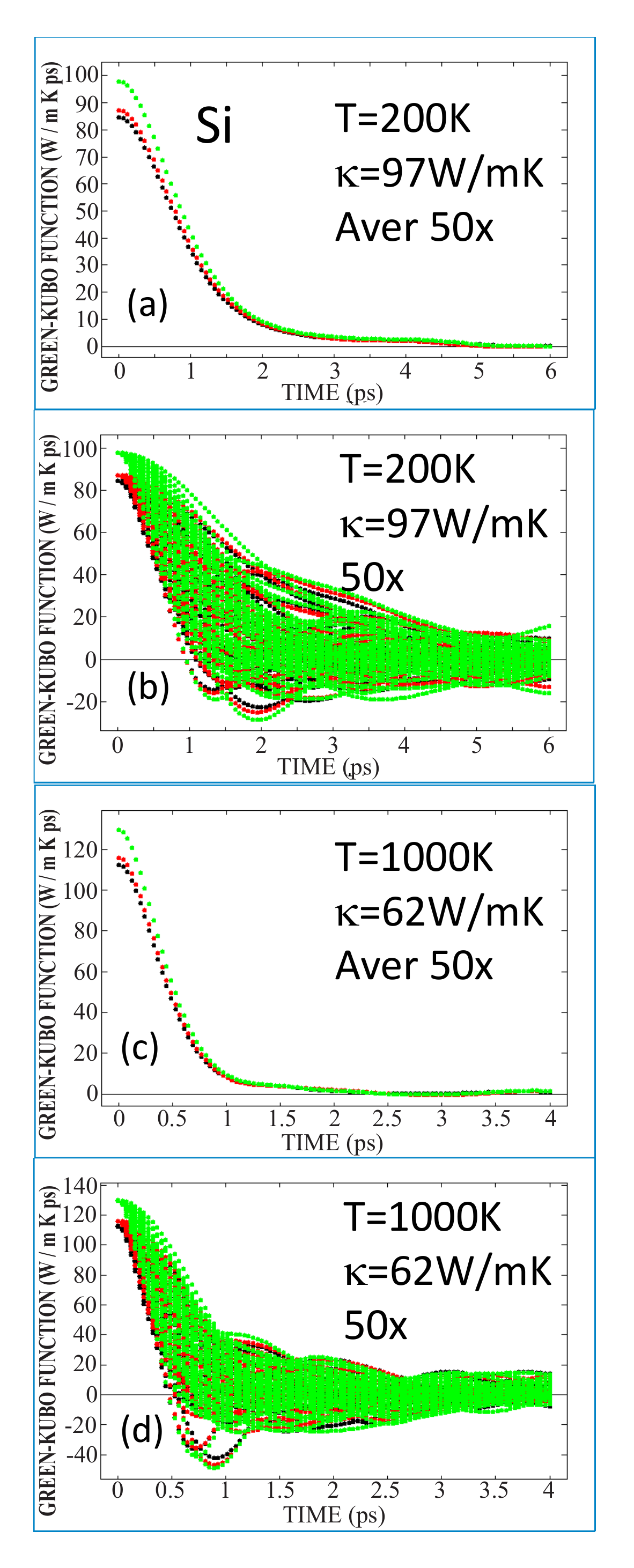} 
\caption{Silicon, Si. Green-Kubo functions, (b,d), 
 Eqs (\ref{Thermaltensor2}) (\ref{GKtwo}),
 and  averaged Green-Kubo functions, (a,c), Eq.(\ref{GKtime}).
For lattice thermal conductivity and 50 DP. 
}
\label{4FigureGKSi4}
\end{figure} 
\begin{figure}[t!]
\includegraphics[width=0.45\textwidth]{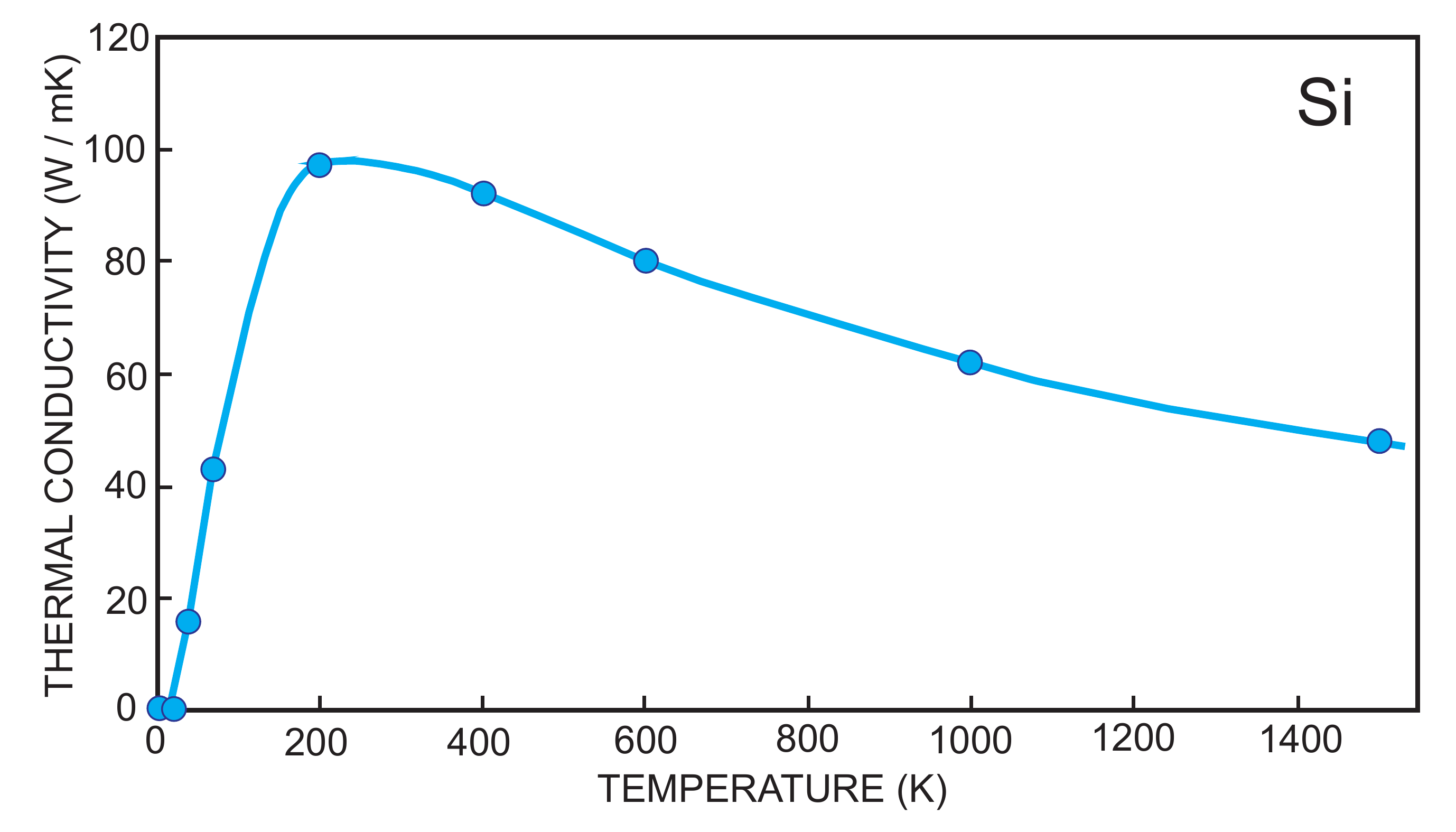}
\caption{Silicon Si. Lattice thermal conductivity, Eq.(\ref{Thermaltensor2})
}
\label{5FigureOpticSi5}
\end{figure}
\subsection{Silicon LTC}
\par
At this point we may proceed a calculation of the LTC for Si using the DP 
determined in previous section.
According to Eq.(\ref{Thermaltensor2}) these DP are used to average 
over $i=1,\dots N_i$  the thermal conductivity.
Phonon frequencies arising from  $N_i$ eigenvalue solutions,
Eq.(\ref{eigenvalue}), can be plotted as phonon dispersion curves. The averages
of the plotted curves reflect the magnitude of anharmonicity in any point of the Brillouin zone.
They are just plotted  on Fig.\ref{3FigureMapSi3}  being results of 500 DP. 
One sees the broadening of anharmonic peaks with rising temperature. 
Notice that the same 500 DP could have been  used to plot phonon dispersion curve, 
density of states, thermodynamic in anharmonic state
and group velocities and LTC.
\par
Using the formulae for thermal conductivity Eq.(\ref{Thermaltensor2}),
the LTC has been calculated
for temperatures $T =  40, 70, 200, 400, 600, 1000, 1500 K$. 
The 50 DP were created for each temperature.
During these procedures the lattice constant of $a=5.3847$ \AA $\,\,$ was recorded, and we
observed that the accompanied  pressure was  stabilized at about $5.6kbar$. 
However, due to somewhat 
lower accuracy of determined  acoustic modes,  the phonons having 
frequency below 0.5THz were removed from contributing to atomic 
configuration, while creating DP patterns. Hence, low frequency phonons 
could be not sufficiently well represented at the low temperature region of thermal conductivity.
\par
Fig.\ref{4FigureGKSi4}b,d shows the detailed behaviour of the Green-Kubo 
functions, Eq.(\ref{GKtwo}),
for $T=200K$ and $T=1000K$ and  
for DP $i=1,\dots 50$ diagonal components 
$(\alpha,\alpha) = (xx, yy, zz)$, plot from bottom to top 
in color order red, black, green.  It is seen that the Green-Kubo functions
are presenting some scatter, which at $t=0$ starts from  value
$\frac{1}{Nr}\sum_{{\bf k},j} Z_{\alpha,\beta}^{(i)}({\bf k},j)$, then after a several $ps$ 
becomes wide and tends at large $t$ to a single line approaching infinity at 
$G_{\alpha,\alpha}^{(i)}(t)$=0.
\par
Fig.\ref{4FigureGKSi4}a,c demonstrates that the global relaxation time for 
$T=200K$ and $T=1000K$ are about $5.43 ps$ and $2.73 ps$, respectively, 
meaning that longer relaxation occurs at lower temperature.
This figure, Fig.\ref{4FigureGKSi4}a,c  shows
the global relaxation times plots, Eq.(\ref{globalrelaxation}), 
being the averaged of Green-Kubo functions,  
$G_{\alpha,\beta}(t)$,  Eq.(\ref{GKtwo}),
shows that the function 
really vanishes at larger time $t$ values. In other words, the time dependent $cosines$ 
periodic functions, 
which depend on different  phonon frequencies,
will progressively overlap all terms so that at long $t$ $G_{\alpha,\beta}(t)=0$.
All diagonal $xx, yy,  zz$ components are computed,  
and the off-diagonal ones $yz, xz, xy$ vanish due crystal symmetry and numerically.
\begin{figure}[t!]
\includegraphics[width=0.45\textwidth]{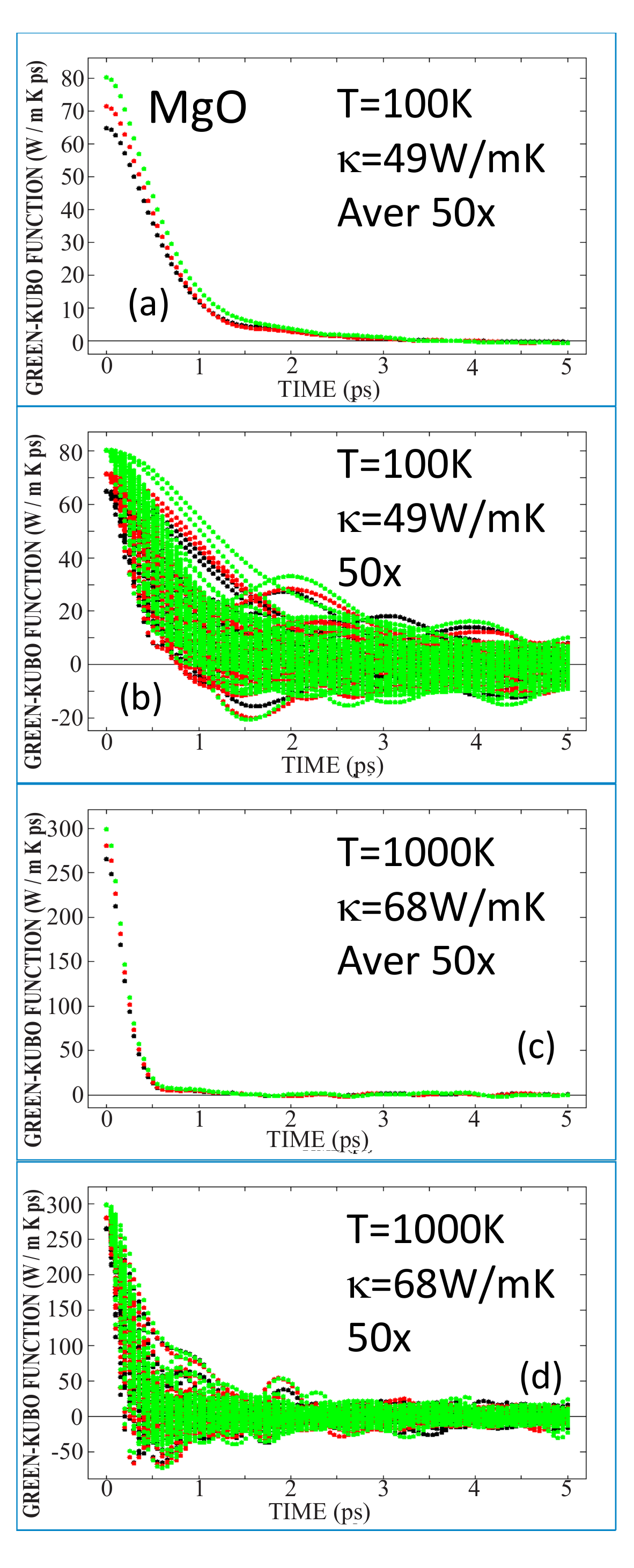}
\caption{Magnesium Oxide, MgO. Green-Kubo functions, (b,d), 
 Eqs (\ref{Thermaltensor2}), (\ref{GKtwo}),
 and  averaged Green-Kubo functions, (a,c), Eq.(\ref{GKtime}).
For lattice thermal conductivity and 50 DP. 
}
\label{6FigureOpticMgO6}
\end{figure} 
\par
Fig.\ref{4FigureGKSi4}b,d shows, that the global relaxation time 
given in Table \ref{kappaphonon} for Si, diminishes with increased $T$.
Moreover, the amplitude function  $Z_{\alpha,\beta}^{(i)}({\bf k},j)$ , Eq.(\ref{Z-function}),
has been obtained in the same 
computational process as the global relaxation time $\tau_{gl}$.
The $\kappa_{\alpha,\beta}$ and $\tau_{gl}$ decrease with increased temperature.
The plot of computation silicon thermal behaviour of LTC is shown on Fig.\ref{5FigureOpticSi5}.
\begin{table}
\caption{Silicon Si and Magnesium Oxide MgO. Calculated averages of the sum
$\frac{1}{3}(\kappa_{1,1} + \kappa_{2,2} + \kappa_{3,3})$
of the lattice thermal conductivities Eq.(\ref{Thermaltensor2}), 
and the global relaxation times, Eq.(\ref{globalrelaxation}).}
\begin{ruledtabular}
\label{kappaphonon}
\begin{tabular}[t]{lccccccc}
Si   $T(K) $ & 40  &  70  &  200 & 400 &  600  &  1000  & 1500  \\ 
\hspace{0.6cm} $\kappa (W/mK)$  &  15.8 & 43.1 &  96.8 & 92.4 &  79.8  &  61.6  &  47.5 \\                                                 
\hspace{0.6cm} $\tau (ps)$ & 5.94 & 5.89 &  5.43 & 4.40 &  3.66 &   2.73 &   2.11 \\
\hline
MgO  $T(K) $ & 20  & 100  &  300 & 600 &  1000  &  1500  &   \\ 
\hspace{0.6cm} $\kappa (W/mK)$  &  3.14 & 49.3 &  110.2 & 92,2 &   68.3  &   50.9  \\                                                 
\hspace{0.6cm} $\tau (ps)$   &  5,56 & 3.63  &  2.67  & 1.82 &  1.28  &  0.88 \\
\end{tabular}
\end{ruledtabular}
\end{table}
\subsection{Magnesium Oxide LTC}
At this point we may proceed a derivation of the LTC for MgO 
with method described above.
Applying the Eq.(\ref{Thermaltensor2}), using data of DP and 
related Hellman-Feynman forces
coming from VASP, we determined                                                                                                                                                                                                                    the LTC coefficients
for temperatures $T =  20, 100, 300, 600, 1000, 1500 K$. 
The 50 DP were created for each mentioned temperature.
The relaxing lattice constant was  $a=4.2462$ \AA$\,\,$.
\begin{figure}[t!]
\includegraphics[width=0.45\textwidth]{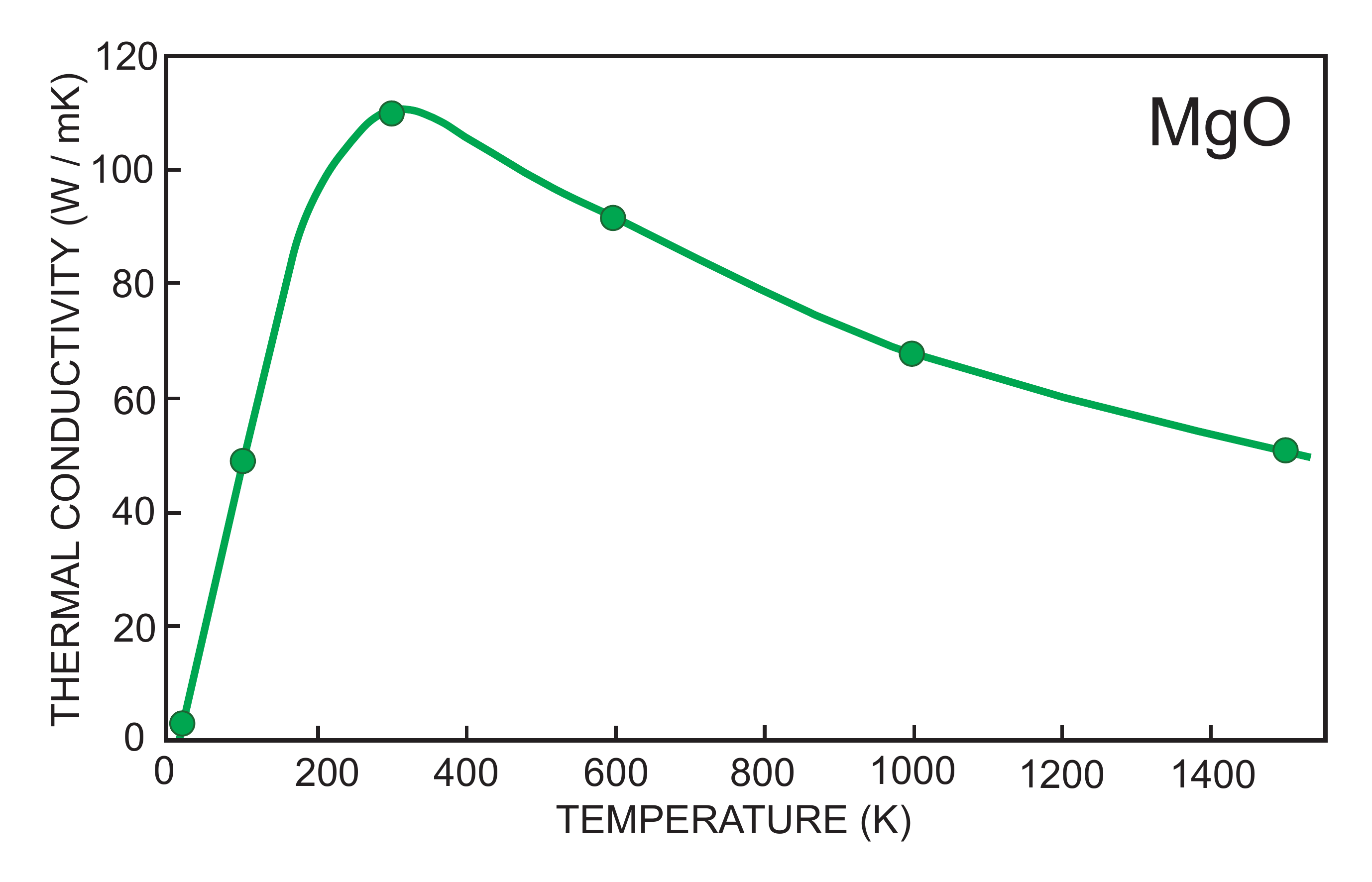}
\caption{Magnesium oxide, MgO. Lattice thermal conductivity,
Eq.(\ref{Thermaltensor2})}
\label{7FigureOpticMgO7}
\end{figure}
The LO/TO splitting effect, present in  MgO, was taken into account substituting
values of ionic 
charges divided by electronic dielectric constant. The optical phonon branches, 
which depend
on effective charges, are located at high frequencies and, as one should expect, have
little influence on the thermal conductivity.
The remaining calculation were performed in analogy to procedures used in Si, 
and described above.
\par
Fig.\ref{6FigureOpticMgO6}b,d shows examples of the detailed behaviour 
of the Green-Kubo functions, 
Eq.(\ref{GKtwo}) for DP $i=1,\dots 50$ for diagonal components 
$(\alpha,\alpha) = (xx, yy, zz)$, plot from bottom to top 
in color order red, black, green.  It is seen that the Green-Kubo functions
are somehow scattered.   At $t=0$ they start from  value
$\frac{1}{Nr}\sum_{{\bf k},j} Z_{\alpha,\beta}^{(i)}({\bf k},j)$, 
then after a time of several $ps$ 
become wide and  tends, at large $t$, to gather to a single line approaching infinity at 
$G_{\alpha,\alpha}^{(i)}(t=0)$=0.
\par
The relaxing lattice constant was  $a=4.2462$ \AA $\,\,$.
The thermal conductivity analysis were run for 
$T=20, 100, 300, 600, 1000, 1500K$.  The structure was stabilized at 
about $10kbar$.
All $N_i=50$ DP were created with removed acoustic phonon modes below $0.5THz$.
There was 192 exact wavevectors with the same exact point as for Si.
The optical phonon branches, which dependent
on effective charges, are located at high frequencies and, as one should expect, have
little influence on the thermal conductivity.
The remaining calculation were performed in analogy to procedures used in Si, 
and described above.
\par
The Figs \ref{6FigureOpticMgO6}b,d and \ref{6FigureOpticMgO6}a,c 
show bundles of Green-Kubo $G_{\alpha,\alpha}^{(i)}(t)$,
and averaged Green-Kubo functions the 
$G_{\alpha,\alpha}(t)$, respectively,  Eq.(\ref{GKtwo}).
The figures for MgO look very similar to respective figures of  Si.
They also present the diagonal elements of the thermal conductivity tensors.
The global relaxation times were estimated to be 3.63 ps and 1.28 ps for 
T=100K, 1000K, respectively.
The temperature behaviour of LTC is shown on Fig.\ref{7FigureOpticMgO7}. 
Similarly to Si, an abrupt
decrease of LTC is observed below $T=200K$.
\section{HIGH  THERMAL CONDUCTIVITY}
\subsection{Elastic Tensor and Equation of Motion}
\par
A material with room-temperature thermal conductivity value larger than $100 \,W/mK$
is regarded as a high thermal conductivity material \cite{srivastava,slack1AlN}. 
Such an effect can be achieved
either by extending the relaxation time $\tau^{(i)}({\bf k},j)$, or/and  increasing of the 
amplitude function value $Z_{\alpha,\beta}^{(i)}({\bf k},j)$,  Eq.(\ref{Z-function}).
Here, we propose to justify the following approach, which might be able to provide
high values of HTC. 
\par
It is well known that material
temperature is mainly governed by its atomic vibrations, which perform  
vibrations with amplitudes of order $0.01 - 0.20$ \AA $\,\,$. 
Such vibrations in form of the phonons occur 
in the whole crystal. Cooling/heating crystal causes to decrease/increase the atomic 
vibration amplitudes.
For HTC, we are tempting to consider mainly the long, even very long  vibrational waves,
which do not care much on the atomistic details of the material structure. Then, the best 
is to use the elastic theory. Within the elastic theory 
it is convenient to determine supercell lattice as element of the crystal space, 
where the elastic waves would propagate. So elastic theory would allow to 
deform the supercell with no need to specified the atoms. In particular, 
one may study properties  through the elastic tensor, 
which  changes with deformations of supercell.
One should, however remember that the elastic tensor is determined by the atomic
interactions and atomic configurations.
\par
There is another reason to pay attention to elastic wave method.
The HTC is determined by very long waves, having length even above  microns.
Such long acoustic phonons are usually represented as acoustic phonons.
The frequencies of such 
acoustic phonons should be of very small and of very  high accuracy, 
what frequently is difficult to achieve
within lattice dynamics method alone, in particular in complex and low symmetry structures.
A selection of elastic waves guaranties the correct
values of elastic wave frequencies in vicinity of the $\Gamma$ point.
Indeed, sometimes in more complex crystals, the long acoustic phonons break the
crystal symmetry due to their vibrations, generally guarantied by 
the translation-rotation invariances 
and dynamically violate the acoustic phonon properties.
Therefore, it seems to be reasonable to 
replace the acoustic phonon modes by the elastic waves.
The elastic waves  diminish these effects. The elastic theory
approach  would operate with waves characterized by
wavevectors of order ${\bf k} = 0.5 - 0.000001$  \AA${\AA}^{-1}$, 
which cover the object sizes from nanometers- to microns.
In this way one also may  introduce the influence of material imperfections 
on the HTC  thermal conductivity. 
\begin{figure}[t!]
\includegraphics[width=0.49\textwidth]{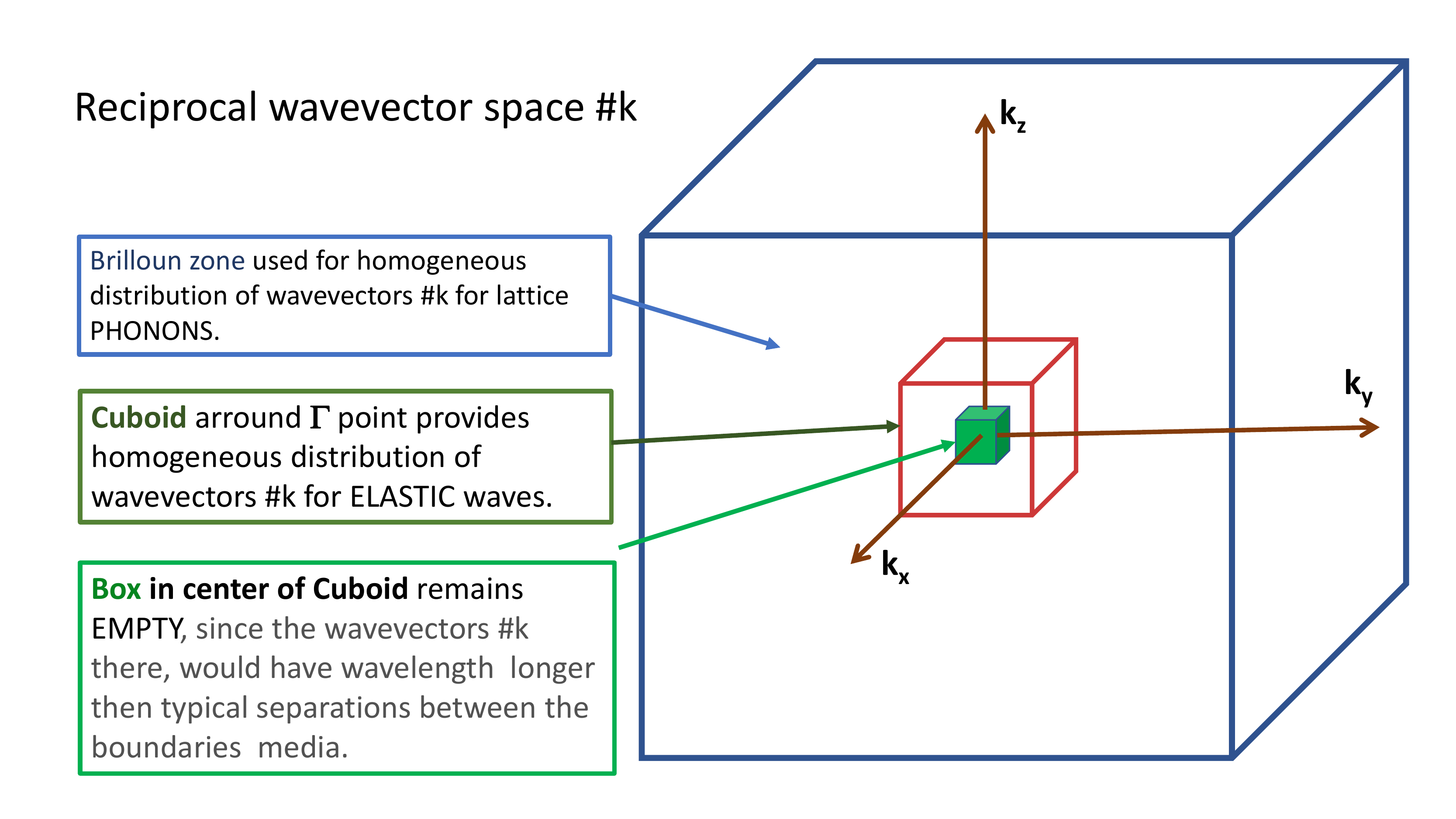}
\caption{(Color online)
The {\it cuboid} volume to which are limited the wavevectors ${\bf K}$ influencing
the deformations of elastic constants tensors due to phonons displaced 
in the supercell by the presence of deformed elastic tensor (DET).
}
\label{8CuboidFig8}
\end{figure}
\par 
The equation of motion for elastic medium is formulated 
for elastic plane waves, which are written as
\begin{equation}
  U({\bf X},t) = A_J exp(\Omega({\bf K},J)t - {\bf K}{\bf X})
\end{equation}
where $U_J({\bf X},t)$    is the supercell material 
deformation at point ${\bf X}$ of the material,
and $J=1,2,3$ is index of elastic mode.
The $A_J$ is a component of vibration amplitude, 
$\Omega({\bf K},J)$ - angular frequency for elastic wave mode, $t$ time, and
$\bf K$ the wavevector of monochromatic  wave.
The  equation of motion of the elastic waves is called {\sc Christoffel} equation 
\cite{christoffel,huntington,boulanger,sluiter,cottenier,klemens,grimvall,fedorov}

\par
The Hooke's low relates the elastic strain $\epsilon_{KL}$ with 
the stress $\sigma_{IJ}$.
\begin{equation}
\sigma_{IJ} = \sum_{K.L=1}^3 C_{IJKL}\,\, \epsilon_{KL}
\label{hooke}
\end{equation}
where $I, J, K, L$ are indices each from 1, 2 up to 3.
The elastic properties of the material are described by a fourth-rank 
tensor   $C_{IJKL}$ with  $3^4=81$ elements. They can be arranged in a $6\times 6$ 
matrix that is symmetric, with elements $C_{\alpha, \beta} = C_{\beta,\alpha}$.
The elastic tensor in the form of $6\times 6$ matrix 
should usually be available from external  program like VASP.
\begin{equation*}
C_{\alpha,\beta} =
\begin{pmatrix} 
  C_{11} &  C_{12} &  C_{13} &  C_{14} &  C_{15} &  C_{16} &  \\
  C_{21} &  C_{22} &  C_{23} &  C_{24} &  C_{25} &  C_{26} &  \\
  C_{31} &  C_{32} &  C_{33} &  C_{34} &  C_{35} &  C_{36} &  \\
  C_{41} &  C_{42} &  C_{43} &  C_{44} &  C_{45} &  C_{46} &  \\
  C_{51} &  C_{52} &  C_{53} &  C_{54} &  C_{55} &  C_{56} &  \\
  C_{61} &  C_{62} &  C_{63} &  C_{64} &  C_{65} &  C_{66} &  \\
\end{pmatrix}
\end{equation*}
The relations between $C_{\alpha, \beta}$ and $C_{JILM}$ are shown in Table \ref{Cab-Cijkl}.
\begin{table}
\caption{Contraction scheme: indices $I,J,K,L$  in $C_{I,J,K,L}$ are replaced by indices
         ${\alpha,\beta}$ in $C_{\alpha,\beta}$. The same rules works in reverse direction.
         Scheme used by Voigt and VASP. The {\sc PhononA} uses VASP notation}
\begin{ruledtabular}
\label{Cab-Cijkl}
\begin{tabular}[t]{lcccccc}
I,J or K,L          & 11 & 22 & 33 & 23/32 & 13/31 & 12/21 \\
\hline
$\alpha$ or $\beta$ &  1 &  2 &  3 &   4   &   5   &   6   \\ 
Voigt               & XX & YY & ZZ & YZ/ZY & XZ/ZX & XY/YX \\                                                 
VASP                & XX & YY & ZZ & XY/YX & YZ/ZY & ZX/XZ \\
\end{tabular}
\end{ruledtabular}
\end{table}
\par
The stiffness tensor $C_{\alpha,\beta}$ not only contains 
information about static materials deformation,
but also about the elastic waves traveling through the material. 
The equation of motion for the elastic waves can be obtained 
from solution of the  Christoffel equation \cite{christoffel}
\begin{eqnarray}
\rho \Omega^2({\bf K},J) \cdot E({\bf K},J) =
\sum_{I,L}  K_I C_{JILM} K_L \cdot E({\bf K},M)
\nonumber \\ 
\label{chriseqn}
\end{eqnarray}
where $\rho$ represents the mass density. 
The solution of the Cristoffel equation  for each wavevector ${\bf K}$ provides three 
solutions corresponding to elastic waves with definite frequencies.
The equation combines the Cristoffel
$3\times 3$ square  matrix ${\bf M}$ with elements
\begin{eqnarray}
{\bf M}_{JM} = \sum_{I,L} K_I  C_{JILM} K_L
\label{Chrisrmatrix}
\end{eqnarray}
Now, Eqs(\ref{chriseqn}), and (\ref{Chrisrmatrix}) form an
eigenvalue problem that can be routinely solved at arbitrary ${\bf K}$.
The result is a set of three frequencies 
$\Omega^2({\bf K},J)$
and polarization vectors
$E({\bf K})$,
 Since \textbf{M} is real and symmetric matrix, the eigenvalues are real and eigenvectors 
 $E({\bf K},J)$ constitute an orthogonal basis. Furthermore, the property that 
 ${\bf M}$ is a symmetric matrix involves that the $\Omega^2({\bf K},J)$ 
 is real and positive. 
\par 
It is convenient to introduce
auxiliary matrix 
\begin{eqnarray}
G_{J,M} = \sum_{I,L} K_I C_{JILM}K_L
\label{Gmatrix}
\end{eqnarray}
This expression represents the core part of the Cristoffel Eq.(\ref{chriseqn}).
The derivatives of $G_{J,M}$ are needed to specify the group velocity. They
could be derived from three  matrices of order $3\times 3$, which are 
wavevector derivatives of matrix  $G_{J,M}$
\begin{eqnarray}
 \Omega_{0}^{2}(K,J) &=& Diag \left[ \sum_{I,L} K_I C_{JILM} K_L \right]
  \nonumber \\
 \frac{\partial \Omega_{0}^{2}}{\partial K_x}(K,J)  &=& Diag \left[ \sum_L (C_{J1LM} K_L + \sum_I K_I C_{JI1M} \right]
  \nonumber \\
 \frac{\partial \Omega_{0}^{2}}{\partial K_y}(K,J)  &=& Diag \left[ \sum_L (C_{J2LM} K_L + \sum_I K_I C_{JI2M} \right]
  \nonumber \\
 \frac{\partial \Omega_{0}^{2}}{\partial K_z}(K,J)  &=& Diag \left[ \sum_L (C_{J3LM} K_L + \sum_I K_I C_{JI3M} \right]
  \label{gr3matrix}
   \nonumber \\ 
\end{eqnarray}%
All matrices in Eqs (\ref{gr3matrix}) can be numerically diagonalized, which is marked by "{\it Diag}".
The inputs are the right hand matrices, while the outputs constitute of following
eigenvalues $\Omega_{0}^{2}$,
$\frac{\partial \Omega_{0}^{2}}{\partial K_x}$, 
$\frac{\partial \Omega_{0}^{2}}{\partial K_y}$ and
$\frac{\partial \Omega_{0}^{2}}{\partial K_z}$. 
The diagonalization of the above matrices gives their 
eigenvalues.
Moreover, one must also diagonalize the matrix $G_{J,M}$.
Ratio of these data divided by $2$ leads to the group velocities of the elastic waves.
Furthermore, one might find this derivative 
differentiating the Cristoffel equation (\ref{chriseqn}). From Eq.(\ref{chriseqn}) one finds
the group velocity of elastic waves
\begin{eqnarray}
 {\bf V}_{gr}({\bf K},J) =    \frac{1}{2\cdot \Omega_{0}({\bf K},J)} 
 \left({\bf i}\frac{\partial  \Omega_{0}^{2}}{\partial K_x},
       {\bf j}\frac{\partial  \Omega_{0}^{2}}{\partial K_y}, 
       {\bf k}\frac{\partial  \Omega_{0}^{2}}{\partial K_z} 
 \right)
\label{elagroupvelocity22}
\end{eqnarray}
where ${\bf i}$, ${\bf j}$, ${\bf k}$ are versors along $x,y,z$ directions.
\par 
Max Born developed in his book \cite{Born} a method which 
correlates the elastic constants with the slopes of acoustic phonon modes
at a particularly small wavevectors. As a matter of fact, the elastic waves have 
been identified as the acoustic waves at small ${\bf K}$. 
Therefore, expression for LTC thermal conductivity 
should hold  for HTC, with only a few differences listed below.
\par
The lattice thermal conductivity calculations introduced above for phonons 
can also be used for  elastic waves. Although their wavelengths
are evidently longer than the size of 
the supercell used in {\textit ab initio} calculations, 
one may account the long wavelength doing the following:
(i) create phonon displacement patterns DP in conventional supercells, for example the same
as for phonons, (ii) see that the displaced atoms also
cause changes of the elastic constants, 
(iii) call {\it deformed elastic tensor} (DET), which conventionally will be the 
supercell  deformed itself, what results in
symmetry lowering to DET.
\par
By solving now the Cristoffel equation, Eq.(\ref{chriseqn}), \cite{parlinski713} 
with DET, one calculates the elastic wave frequencies, 
finding their frequency changes with respect of  ideal  supercell tensor.
One may say that the Cristoffel equation rebuilds the ideal elastic wave to whole space 
from the crystal segment belonging to DET limited to studied supercell.
Moreover, the group velocities Eq.(\ref{elagroupvelocity22}), can also change.
These changes influence the relaxation time of conducting objects acting in thermal conductivity. 
\par
The relation for high thermal conductivity (HTC) 
formulated in analogy with lattice thermal conductivity LTC, Eq.(\ref{Thermaltensor2}),
for the simulation with the deformed elastic tensor DET  reads
\begin{widetext}
\begin{eqnarray}
\nonumber \\
\kappa_{\alpha,\beta}^{HTC} &=& \frac{\hbar ^2}{NrV_{puc}k_BT^2}
\frac{1}{N_{i}} \sum_{i=1}^{N_i}
\int _0^\infty dt    \sum_{{\bf K},J}^{Cuboid} (\Omega^{(i)}({\bf K},J))^2
{\bf V_{puc}}^{(i)\alpha}_{gr}({\bf K},J) {\bf V}^{(i)\beta}_{gr}({\bf K},J) 
\nonumber \\
&\times&
(n^{(i)}({\bf K},J) +1)) (n^{(i)}({\bf K},J) 
\times \frac{1}{2}
 \,\, cos[2(\Omega^{(i)}({\bf K},J) - \Omega^{(0)}({\bf K},J))t]
\nonumber \\
\label{Thermaltensor3}
\end{eqnarray}
\end{widetext}
where $N_i$ is the number of DET - deformed elastic tensors, 
$V_{puc}$ volume of the primitive unit cell. 
The {\bf\it Cuboid}, see Fig. \ref{8CuboidFig8} is a volume in the reciprocal zone, 
with  center point at  $\bf{K}=0$.
In LTC Eq.(\ref{Thermaltensor2}), the summations $\sum_{\bf k,j}$ 
run homogeneously over the whole Brillouin zone.
In HTC  Eq.(\ref{Thermaltensor3}), the summations $\sum_{\bf {\bf K},J}$ 
should run over small wavevector volume around ${\bf K}=0$ as  
indicated in Cuboid, $\sum_{{\bf K},J}^{Cuboid}$.
The central green box should always remain empty 
(it is direct space beyond volume of the sample), 
so there no wavevectors should be positioned.
In the volume of the larger inner box (between green and red boxes),  the wavevectors 
$\bf{K}$ for elastic waves should be placed. 
Notice, that a lot of wavevectors can determine crystal phonons, 
much less  wavevectors are indexing the elastic waves.
The central box excludes such long wavevectors $\bf{K}$, which 
surpass the sample macroscopic size, or  
the mean distance between boundaries existing in the media.
In general, the wavevectors should be places at random, unless the wavevectors amplitudes
and positions express some superstructure being a new object of the study. Then, the positions 
and amplitudes of wavevectors  $\bf{K}$ could be derived from the object in the direct space and
then transformed to the cuboid by  three-dimensional Fourier 
transform. Of course, the shape of the Cuboid may change to adapt to the studied object.
Remember that sets of wavevectors for phonons  $\bf{k}$    and elastic waves $\bf{K}$
are needed for LTC Eq.(\ref{Thermaltensor2})  
and HTC  Eq.(\ref{Thermaltensor3}) expressions, respectively. And that the HTC must 
be described by wavevectors from the elastic wave region.
\begin{figure*}[t!]
\includegraphics[width=0.98\textwidth]{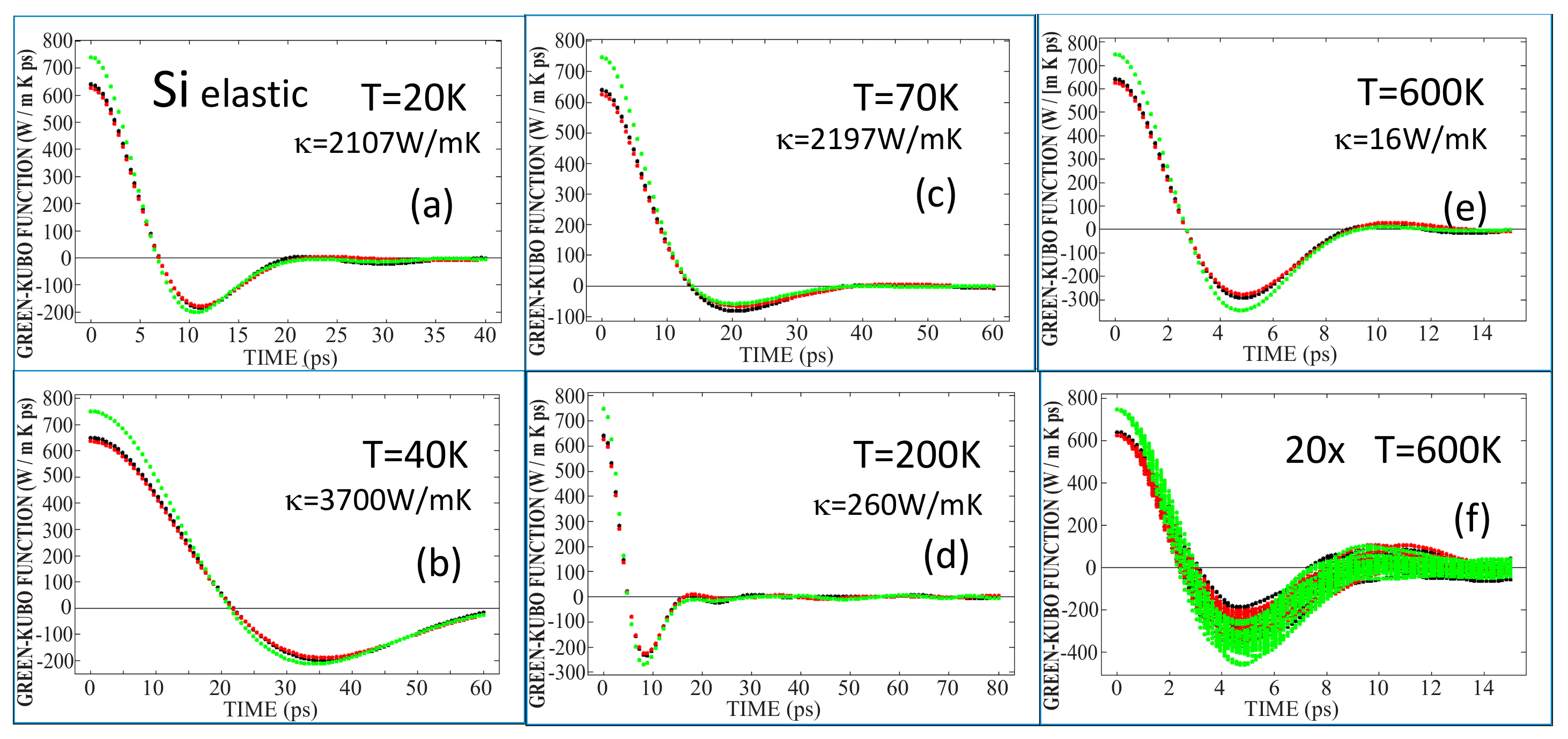}
\caption{Silicon Si. (a,e) HTC averaged Green-Kubo functions calculated from
Eqs(\ref{Thermaltensor3}) and (\ref{GKtwo}), 
(f) Green-Kubo functions for HTC (elastic waves) calculated directly 
from Eq.(\ref{Thermaltensor3}). 
Each run used 20 DET. 
}
\label{9FigureHighGKSi9}
\end{figure*} 
\subsection{ Silicon  HTC}
\par
To calculate the HTC of silicon the formulae for thermal conductivity 
Eq.(\ref{Thermaltensor3}) was used. The 20 DP were prepared for each
$T = 7, 20, 40, 70, 200$ and $600K$, and  then 20 DET's were created in each case.
During heating the lattice constants and pressure stayed constant as observed for LTC of Si.
The elastic tensors, called also elastic modulus, were calculated on {\sc VASP}, \cite{vasp,saxe}.
One must also add that the cpu calculation time  of this process is long in comparison
to cpu run for elastic tensor possessing some symmetry elements. It is a result of the fact that 
the DET's do not have any symmetry, hence, it requires to calculate much more iterations.
\begin{figure}[t!]
\includegraphics[width=0.49\textwidth]{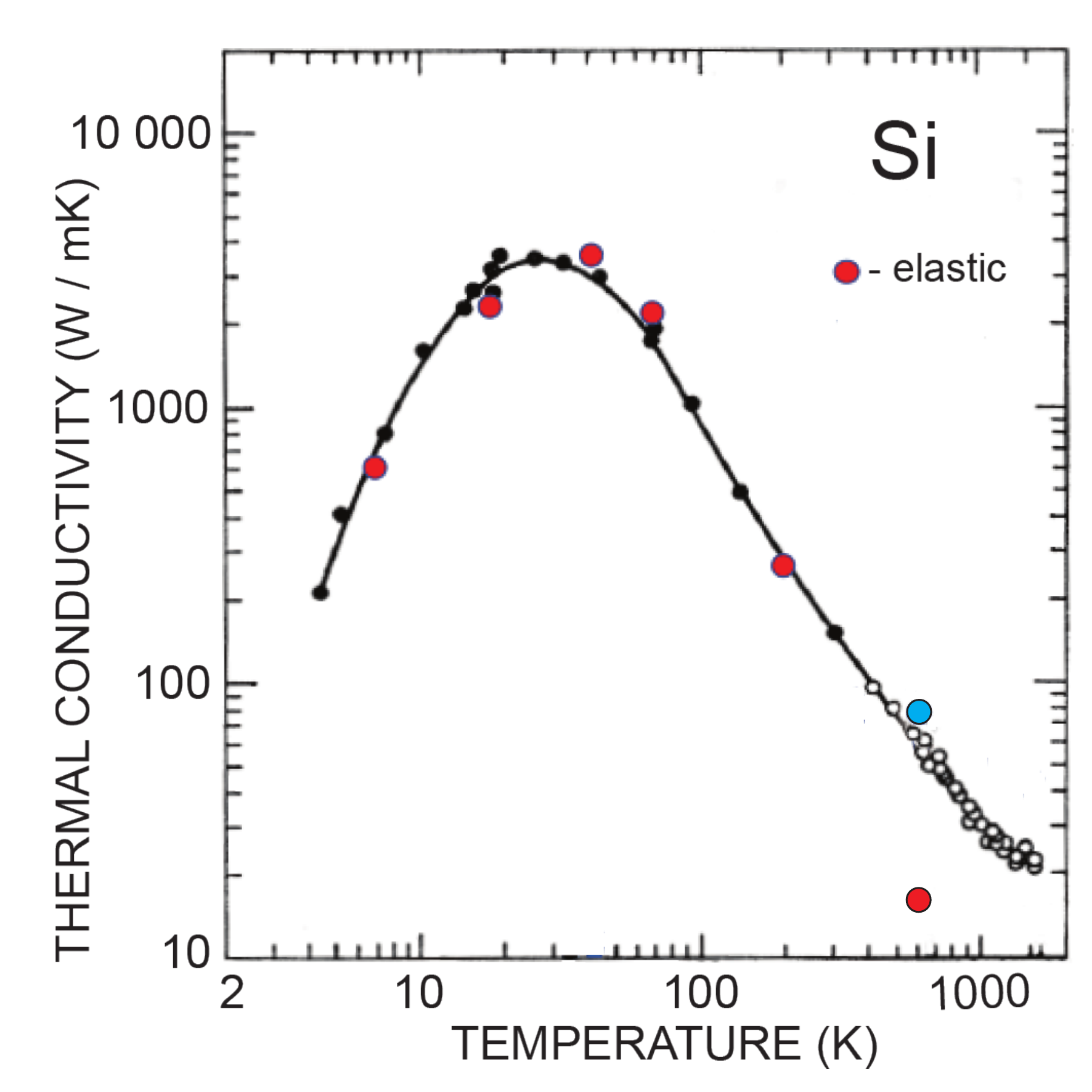}
\caption{Silicon Si. (red) Calculated HTC data using Eq.(\ref{Thermaltensor3})
making use of elastic waves.
(blue) Single point of LTC data at $T=600K$,  imported from Fig. \ref{5FigureOpticSi5}.
(black) Measured thermal conductivity from  Ref. \cite{slack3SiGe}. 
}
\label{10FigureHighSi10}
\end{figure}
\par 
Moving back to HTC, one should supply some information on the 
{\textit crystal microstructure}, and include it to Eq.(\ref{Thermaltensor3}).               
The minimum information should indicate the possible range of the wavelengths of the 
elastic waves before they reach the boundaries, which hinder their travel and then  determine
the expected HTC.
In the present stage of the current theory we may propose to select the proper wavelengths, 
or rather wavevectors 
of the elastic waves only, and check whether they lead to correct results 
observed in experiments. The idea is that the shortest  wavelengths 
$\lambda_{min} = 2\pi/{\bf K}_{max}$
of the elastic waves start from a distance just above the active
wavelength of low frequency of acoustic phonons, and
spreads to longest distances to ''boundaries'' $\lambda_{max} = 2\pi/ {\bf K}_{min}$. 
It is expected that elastic waves characterized by wavevectors ${\bf K}$ from the interval 
${\bf K}_{min} < {\bf K} < {\bf K}_{min}$ could propagate in the crystal without obstacles.
Generally, such precise information is missing. Moreover, the crystal microstructure may depend
on distributions of the boundaries within the crystal, microcracks,
kind of defects and impurities, etc. and as such they ought to be a topic of separate study. 
Here, the volume of the perfect crystal is  represented by a 
cuboid, Fig. \ref{8CuboidFig8} filled with the wavevectors $K$ 
with an exception of the inner box, which should be empty. In the direct space 
the empty box represents the outer part beyond the crystal and   the cuboid   represents  
surface layers of the real crystal, expected to suppress propagation of elastic waves.   
\begin{figure*}[t!]
\includegraphics[width=0.98\textwidth]{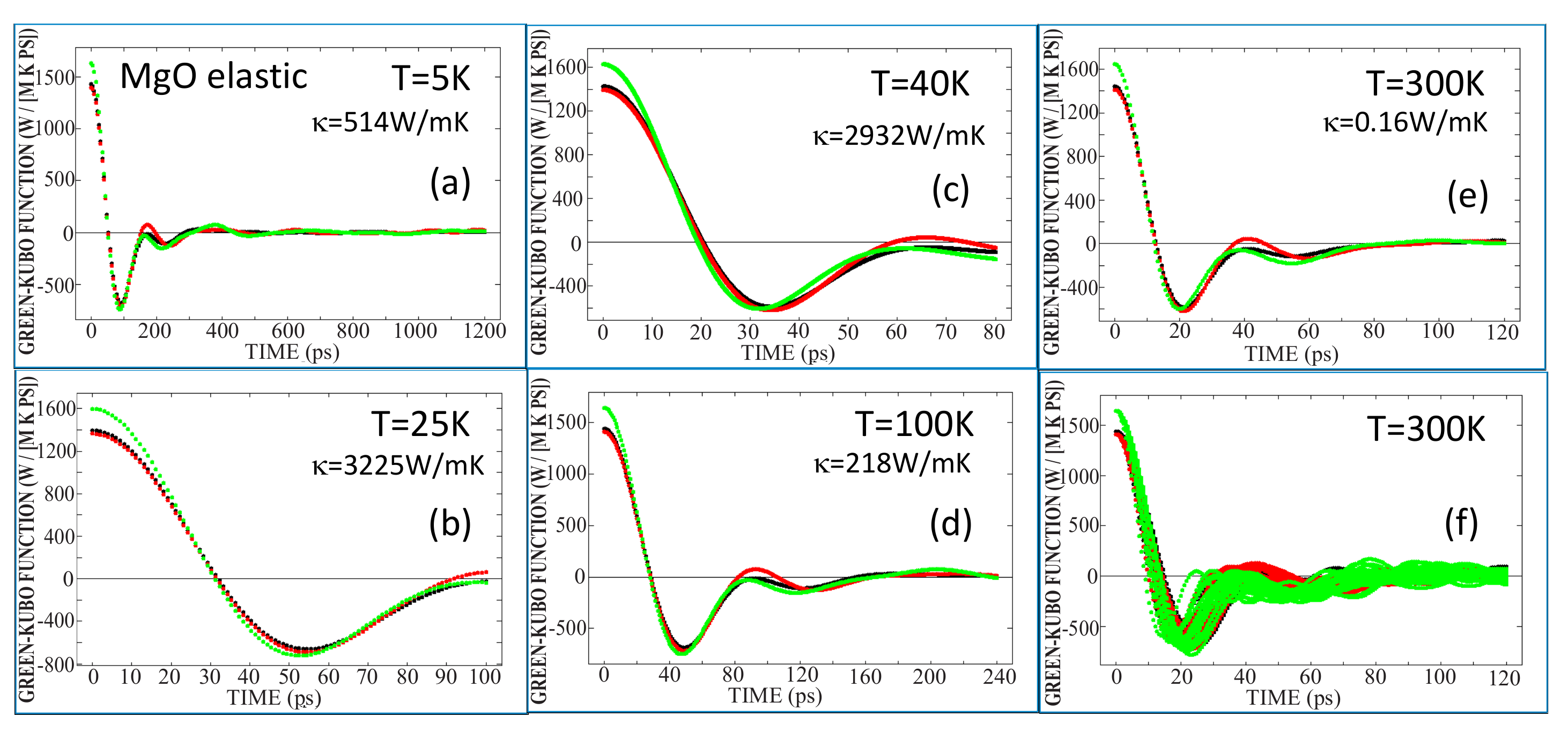}
\caption{Magnesium oxide, MgO. (a,e) HTC averaged Green-Kubo functions calculated from
Eqs(\ref{Thermaltensor3}) and (\ref{GKtwo}), 
(f) Green-Kubo functions for HTC (elastic waves) calculated directly 
from Eq.(\ref{Thermaltensor3}). 
Each run used 20 DET. 
}
\label{11FigureHighGKSi11}
\end{figure*} 
\par
In the present stage of the theory the microstructure information could be accounted
in a primitive way, namely by limiting the boundary to select the wavevectors 
$\bf K$ belonging to cuboid  in the summation $\sum_{{\bf K},J}^{Cuboid}$ of the formula
Eq.(\ref{Thermaltensor3}). For the present study of Si, the random wavevectors were selected out the 
volume of the cuboid confined  by minimal value
${\bf K}_{min}= 0.00001$\AA ${\AA}^{-1}$ 
to maximal value of
${\bf K}_{max} = 0.030$\AA ${\AA}^{-1}$
The inner box inside ${\bf K}_{min} = 0.00001$ \AA ${\AA}^{-1}$
was left empty.
The wavevectors ${\bf K}_{max}$ and ${\bf K}_{min}$ correspond to the lengths of elastic waves
from $0.021\mu$m to $63\mu$m, respectively. Since in the cuboid the wavevectors distribution is 
homogeneous the amount of elastic waves taken in derivation of the HTC 
close to $63\mu$m is less than in vicinity  of $0.021\mu$m.
 
\par
The above mentioned summation in the cuboid was used to
compute the Green-Kubo functions. The selected function are shown on Fig.\ref{9FigureHighGKSi9}.
The vertical axes of plots are mainly determined by the group velocities 
and temperature occupation distributions and  they stay almost constant. But the horizontal axes cover
changeable time intervals. Thus, time seems to decide about the magnitude of the relaxation times and
thermal conductivities.
In particular after initial maximum at $t=0$, the Green-Kubo function diminishes to negative minimum.
Such a decrease is caused by the difference of two cosines occurring in 
the relaxation time expression, Eq.(\ref{Thermaltensor3}). Namely, small differences  between the 
elastic wave frequencies of deformed and perfect supercells lead to  minimum at longer times,
in contrary to opposite situation with large frequency difference and the minima occurring
at shorter times. The  Green-Kubo functions accompanied the elastic waves can be seen 
on the plots. The minima of the (a-f) plots are drawn from 20 runs of DET's. 
On (f) all 20 plot are seen
and some smearing of the data are observed.
\begin{figure}[t!]
\includegraphics[width=0.49\textwidth]{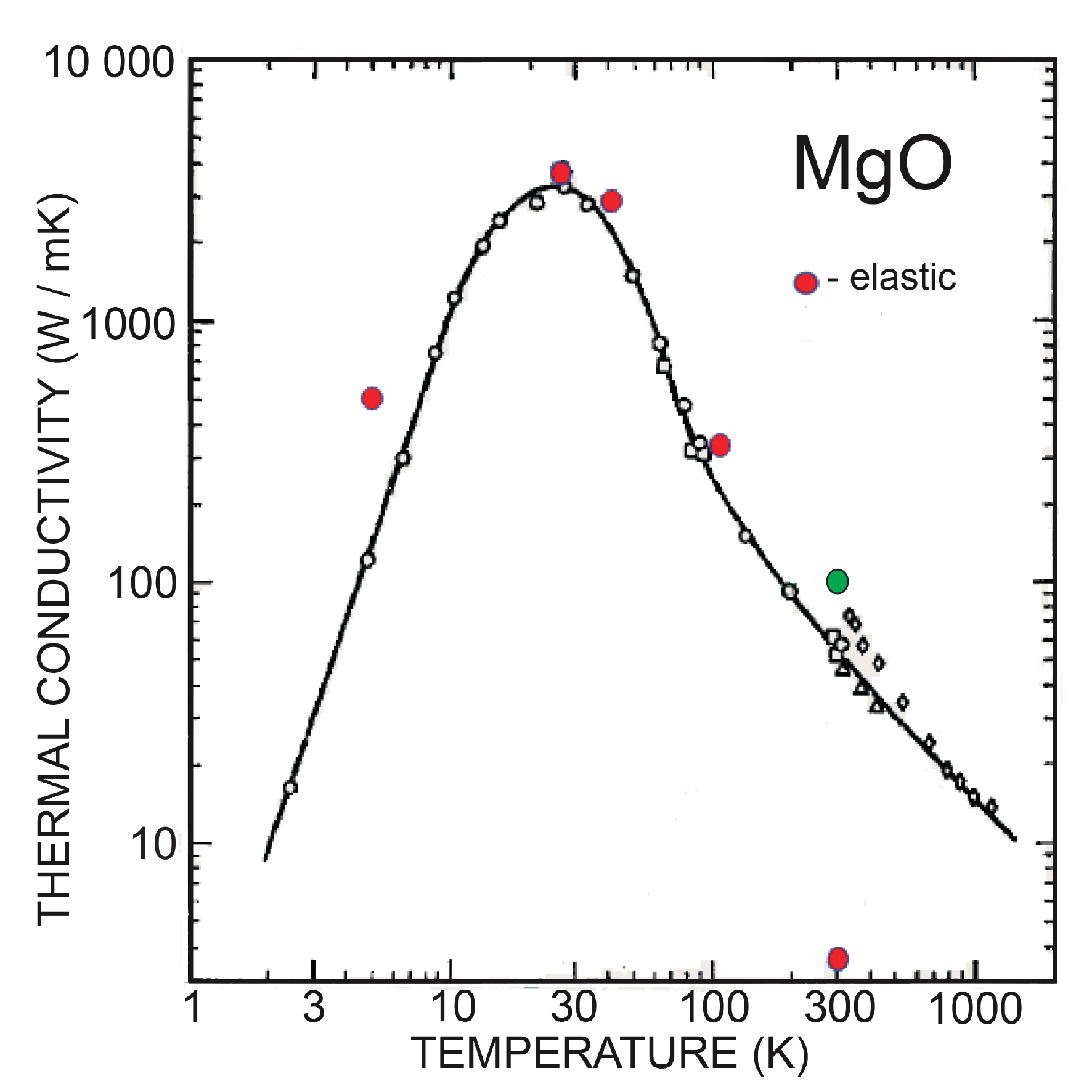}
\caption{Magnesium oxide, MgO. (red) Contributions of  only elastic waves 
to high thermal conductivity. (green) Single point of LTC data at $T=300K$, 
 imported from Fig.\ref{7FigureOpticMgO7}.
(black) Experimental points from \cite{slackMgO} 
}
\label{12FigureHighSi12}
\end{figure}
\par
Relation Eq.(\ref{Thermaltensor3}) gives also the numerical value of thermal conductivity 
as a function of temperature. For silicon the low temperature HTC results are plotted on  
Fig.\ref{10FigureHighSi10}   The LTC data reaches maximum $96.8 \, W/mK$, while the HTC is elevated 
to $3700 \, W/mK$. This high difference is mainly caused by the increase of the global 
relaxation times in HTC mechanism.
Longer relaxation times result in higher values of $\kappa_{\alpha,\beta}$. 
The silicon single crystal, for the HTC measurements
by Glassbrenner and Slack reported in 
\cite{slack3SiGe} and used for the low-temperature
measurements, was grown from high-purity silicon.
The growth process was made with care in order to make the crystal to be oxygen and dislocation free.
Then, any vacancy clusters were less than one micron in diameter. 
The sample boundary sizes were not reported.
\par
It is worth to mention that at low temperature the occupation factor   
$(\frac{\omega}{T})^2 (n(\frac{\omega}{T}) +1) n(\frac{\omega}{T})$ in
Eq.(\ref{Thermaltensor3})  reduces the intensity of the Green-Kubo function.
These properties also force to decrease
HTC close to $T=0K$ to a very small value. 
With increasing temperature the mentioned thermal factors 
approach $1$. Similar effect has been seen for MgO.
\begin{table}
\caption{Silicon Si and Magnesium Oxide MgO. Calculated  averaged sums
$\frac{1}{3}(\kappa_{1,1} + \kappa_{2,2} + \kappa_{3,3})$
of the high thermal conductivities Eq.(\ref{Thermaltensor3}), 
and the global relaxation times, Eq.(\ref{globalrelaxation}).}
\begin{ruledtabular}
\label{kappaelastic}
\begin{tabular}[t]{lcccccc}
Si   $T(K) $ &  7  &  20  &  40  & 70 & 200 & 600   \\ 
\hspace{0.6cm} $\kappa (W/mK)$  & 628 & 2107 & 3700 & 2197 & 260  &  21.2\\                                                 
\hspace{0.6cm} $\tau (ps)$ &  29.6 & 21.6 & 20.7  & 10.4 & 1.63 &  0.11 \\
\hline
MgO  $T(K) $ & 5 &  25 &   40 &   100 &  300   &   \\ 
\hspace{0.6cm} $\kappa (W/mK)$  &  514 & 3225 & 2932 & 218  & 0.14 &  \\                                                 
\hspace{0.6cm} $\tau (ps)$      & 1.15 &  9.94 & 8.81 & 8.36 & 0.13 &                                                                              \\
\end{tabular}
\end{ruledtabular}
\end{table}
\subsection{ Magnesium Oxide  HTC}
\par
To calculate the HTC of magnesium oxide, MgO atomic displacements DP were prepared for each
$T = 5, 25, 40, 100$ and $300K$ and next 20 DET's tensors were created for each $T$ using {\sc VASP}.
During heating the lattice constants and pressure behaved as observed for LTC of MgO.
\par
The below mentioned summation within the cuboid was used to
compute the Green-Kubo functions. The selected functions are shown on  Fig.\ref{11FigureHighGKSi11}.
The vertical plot axes are mainly determined by the group velocities 
and temperature occupation distributions and  they stay almost constant. But the horizontal axes cover
changeable time intervals. Thus, time seems to decide about the magnitude of the relaxation times and
thermal conductivities.
\par 
To calculate HTC of MgO one should select data for the cuboid. The following wavevectors 
have been proposed: 
maximal values of
${\bf K}_{max}= 0.0030$ \AA ${\AA}^{-1}$ 
 and minimal value of
${\bf K}_{min} = 0.00001$ \AA ${\AA}^{-1}$. 
The selected cuboid wavevectors for MgO
correspond to boundaries of the elastic waves being in the range
from  $0.21\mu$m to about $63\mu$m, respectively.
This information was used in Eq.(\ref{Thermaltensor3}) to  compute 
the time dependent Green-Kubo HTC functions, and later 
plotted on  Fig. \ref{11FigureHighGKSi11}.
The HTC data reaches maximum $3700 \, W/mK$,
The variation of the HTC in MgO are presented on Fig.\ref{12FigureHighSi12}.
The $T = 5, 25, 40, 100K$ fit to measured data, but the contribution 
of HTC at $T=300K$ practically vanishes. At $T=300 K$ only LTC contributes.
\par
The two approaches provide thermal conductivity  LTC and HTC.
Therefore, it rises a question what happens 
in the temperature interval between the LTC and HTC regions.
The present results give 
for $Si$ at $T=600K$: HTC: $21,2 W/mK$, LTC: $84 W/mK$   Exp:$ 84.5 W/mK$, and  
for $MgO$ at $T=300K$: HTC: $0.14 W/mK$, LTC: $110 W/mK$   Exp: $75 W/mK$.
This  means that LTC mechanism is used at higher $T$ 
and then at lower $T$ it became replaced by HTC processes. 
Probably, this effect occurs 
in special materials only.
\subsection{Microstructure in HTC}
\par
The current approach relates the HTC and microstructure 
in the region of  low temperatures.
The conventional process to carry on the HTC calculations would be to
look into cuboid for the wavevectors data 
to get  an agreement between  calculated 
and measured 
$\kappa_{\alpha,\beta}^{HTC}$. 
It would mean to find right values 
of the   wavevectors inserted to cuboid.
However, the reverse process would be more valuable. 
First the microstructure features are foreseen
and used to modify the cuboid, and next to collate 
the calculated data with the behaviour of the measured HTC. 
In this case the method could have some  predictive power, which might help 
to design the required properties of the material, say microstructure. 
\par
As a test we have  inserted to the cuboid of Si crystal being at $T=40K$,
the wavevectors of ${\bf K}_{min}= 0.00001$ \AA ${\AA}^{-1}$  
and ${\bf K}_{max} = 0.00015$\AA ${\AA}^{-1}$,
corresponding to Si  boundary distances 
from 63 and 4.2 $\mu$m, respectively.
For these wavevectors  HTC  tremendously increases 
the global relaxation time to 1700 $ps$ and
thermal conductivity to $\kappa_{\alpha,\beta}^{HTC} = 320 000 W/mK$.
Other obstacles might diminish/change this value.
\par
Presently the cuboid volume is filled with wavevectors of the same amplitude.
However, having a concept of the microstructure of considered crystal 
one might convert this information
 to the amplitudes of wavevectors placed in the cuboid. Such a project 
is still waiting for  realization. 
\par
There are many studies, which need to combine the microstructure of 
the sample with its thermal conductivity. Here, follows some  example
(1) The thermal energy transport in actinide oxide nuclear fuel materials \cite{Hurley},
thorium dioxide and uranium dioxide. The first has a characteristic maximum 
of thermal conductivity below 40 K. The second has a reduced thermal conductivity, 
in spite of similar crystal symmetries. It happens due to
the presence of elastic phase transition.
(2) The dislocation impact on thermal conductivity \cite{Cheng}. Dislocations induce 
the stress field, which might lead to anisotropy of thermal transport. 
Such a contribution can be estimated. Another goal would be to analyze 
the collection of dislocations on  thermal conductivity.
(3) Thermal properties of the superelastic, which consist of many crystal variant
of shape memory alloys (as NiTi). 
The complex microstructure exists due to well-known 
compatible equations for pair of crystals variants, 
which require to identify the interfaces \cite{MinJyunLai}.

The MD simulations of simple crystal models 
have shown that  realistic microstructures of 
$YBa_2Cu_3O_7$ superconductor \cite{parlinski1995} and 
$LaNbO4$ ferroelastic \cite{parlinski1997} could be obtained
starting from a simple 
crystal model at relatively high temperature, 
and next quenching.
The simulated miscrostructures and those obtained 
from TEM observations are very similar.
It is also worthwhile to mention the effort to increase effectiveness
of thermoelectric  $CaCd_2Sb_2$ \cite{thermoelectrics}. It  has been proposed
to replace $Cd$ by $Mg$, as point defect, to induce significantly phonon scattering,
but maintaining the carrier concentration, lower essentially their LTC 
and increase figure of merit $ZT$.  
\section{CONCLUSIONS AND DISCUSSIONS}
The mechanisms of thermal conductivities  discussed in this article
is based on the anharmonic phonons formalism  handled within 
the non-perturbative approach for crystals, \cite{parlinski} .
Due to it, the formulated theories of LTC and HTC have been retrieved 
starting from different and not conventional anharmonic approaches.
In the LTC case phonons play a role of the  heat transport media
realized by anharmonic phonons of the crystals. 
The anharmonic vibrations of atoms determine the
crystals temperature. Next, one  prepares several displacement pattern DP of atoms
in the crystal supercells, 
with amplitudes  displaced corresponding to studied temperature. 
Then, the forces induced by the displaced atoms are calculated with 
the \textit{ab initio} software,
which permits to solve the set of lattice dynamics equations,
and find information about the harmonic and anharmonic interatomic potentials. 
This technique has been successfully used to create positions, shifts, widths and shapes of anharmonic
peaks and determine the analytical expression for the mode relaxation times for 
all anharmonic modes without  performing 
expansion of the interaction potential over anharmonic terms
and without using the Boltzman equation.
Specially, the relaxation times  could  be calculated analytically,
and this  process needs only to know the differences of anharmonic  
and harmonic frequencies for each segment of the anharmonic phonon mode
belonging to the same $({\bf k},j)$ phonon mode.
\par 
Some crystals require to take into account also the elastic waves, which need to be
considered within different mathematical formalism. The elastic waves, 
travel in the crystal  and
form a strain variation of predefined units of crystal, usually supercells. 
Atomic displacement patterns, similar to those of phonons, create some 
strains, deform supercells, and  hence create the elastic waves. 
Finally, the elastic waves can be found by solving the Cristoffel equation, 
being entirely defined by the elastic constant tensors.
\par
We have shown that the crystal thermal conductivity is determined 
by the Green-Kubo relationship being the 
correlation function of the heat flux. The high thermal conductivity can be calculated from 
products of elastic wave frequencies, elastic wave group velocities, and phonon relaxation times
specific for elastic waves.
\par
Second essential difference between lattice and high thermal conductivity is 
related with the wavevectors summation within the correlation functions. 
In the phonon part all wavevectors need to be used in the sum over the Brillouin zone.
The elastic waves are characterized by very long wavelengths, so only the short 
wavevectors around $k=0$ and lower then $1-4 THz$ participate in the thermal conductivity. 
This effect is applied 
to fix the longest wavevectors as being able to reach 
the sample sizes, or other obstacles which limit the transport of the elastic waves. 
This criterion leads to statement that the calculated  
HTC may agree  with the measurements.
To facilitate the procedure to limit the used wavevectors for the elastic wavevectors, 
the cuboid of $3d$ figure was introduced. In future the cuboid might also allow to study 
thermal conductivity of crystals with defects, microstructure, etc.
\section*{ACKNOWLEDGMENTS}
 The author would like to acknowledge
 Dr Erich Wimmer and Dr Walter Wolf from Materials Design Inc for 
 suggestions and fruitful discussions.
\section*{AUTHOR DECLARATIONS}
  \textbf{Conflict of Interests}\\  
   The author has no conflict to disclose. 
\section*{DATA AVAILABILITY} 
   The data that support the findings of this study are available within the article.
\section*{RFERENCES}
\par

\end{document}